\documentclass[hyper,letterpaper,12pt]{article}
\usepackage{amsmath}
\usepackage[
      colorlinks=true,
      linkcolor=blue,
      urlcolor=blue,
      filecolor=black,
      citecolor=blue,
            ]{hyperref}
\usepackage{color}
\usepackage{graphicx}
\usepackage{amsfonts}
\usepackage{amssymb}
\usepackage{hyperref}
\usepackage{simplewick}

\usepackage{dsfont}

\graphicspath{{./Artwork/}}

\def\vev#1{\left\langle #1 \right\rangle}
\def\W #1{\widetilde{#1}}

\begin{document}

\title{\bf \Large On-shell Methods for Form Factors in ${\cal N}=4$ SYM and Their Applications}

\author
{\large
Gang Yang\footnote{E-mail: yangg@itp.ac.cn}\\
\\ 
\small \emph{$^1$CAS Key Laboratory of Theoretical Physics, Institute of Theoretical Physics,}\\
\small \emph{Chinese Academy of Sciences, Beijing 100190, China}\\
\small \emph{$^2$School of Physical Sciences, University of Chinese Academy of Sciences,} \\
\small \emph{No. 19A Yuquan Road, Beijing 100049, China}
}
\date{}
\maketitle

\begin{abstract}
\normalsize 
\noindent
Form factors are quantities that involve both asymptotic on-shell states and gauge invariant operators. They provide a natural bridge between on-shell amplitudes and off-shell correlation functions of operators, thus  allowing us to use modern on-shell amplitude techniques to probe into the off-shell side of quantum field theory. In particular, form factors have been successfully used in computing the cusp (soft) anomalous dimensions and anomalous dimensions of general local operators. This review is intended to provide a pedagogical introduction to some of these developments. We will first review some amplitudes background using four-point amplitudes as main examples. Then we generalize these techniques to form factors, including (1) tree-level form factors, (2) Sudakov form factor and infrared singularities, and (3) form factors of general operators and their anomalous dimensions. Although most examples we consider are in ${\cal N}=4$ super-Yang-Mill theory, the on-shell methods are universal and are expected to be applicable to general gauge theories.
\end{abstract}

% Dedicated to Xbb and Xxhh 

\newpage
\tableofcontents

%%%%%%%%%%%%%%%%%%%%%%%%%%%%%%%%%%%%
%%%%%%%%%%%%%%%%%%%%%%%%%%%%%%%%%%%%
\section{ Introduction}
\label{sec:intro}

 Significant progress has been made in computing scattering amplitudes in the past thirty years, for which there have been many excellent reviews, see e.g. \cite{Mangano:1990by, Dixon:1996wi, Bern:2007dw, 2011JPhA44a0101R, Dixon:2013uaa, Elvang:2013cua, Henn:2014yza, Feng:2011np, Weinzierl:2016bus, Alday:2008yw}. 
It should be fair to say that among these developments, one of the most important ideas is the use of on-shell methods, such as the spinor helicity formalism \cite{Xu:1986xb, DeCausmaecker:1981jtq, Berends:1981rb, Kleiss:1985yh}, the tree-level recursion relations \cite{Britto:2004ap, Britto:2005fq} and the (generalized) unitarity methods \cite{Bern:1994zx, Bern:1994cg, Britto:2004nc}.
While scattering amplitudes are central physical quantities in quantum field theory, there are other important objects, such as gauge invariant operators. Computing their anomalous dimensions and correlation functions has also been an important subject. 
A question that one may ask is: can the modern advances of scattering amplitudes be applied to more general observables such as anomalous dimensions and correlation functions? 
At first sight the answer seems to be negative, because unlike amplitudes, gauge invariant operators are off-shell, therefore, the on-shell methods seem not applicable. 
Fortunately, this problem can be overcome with the help of form factors.

\begin{figure}[t]
\centerline{\includegraphics[height=5.5cm]{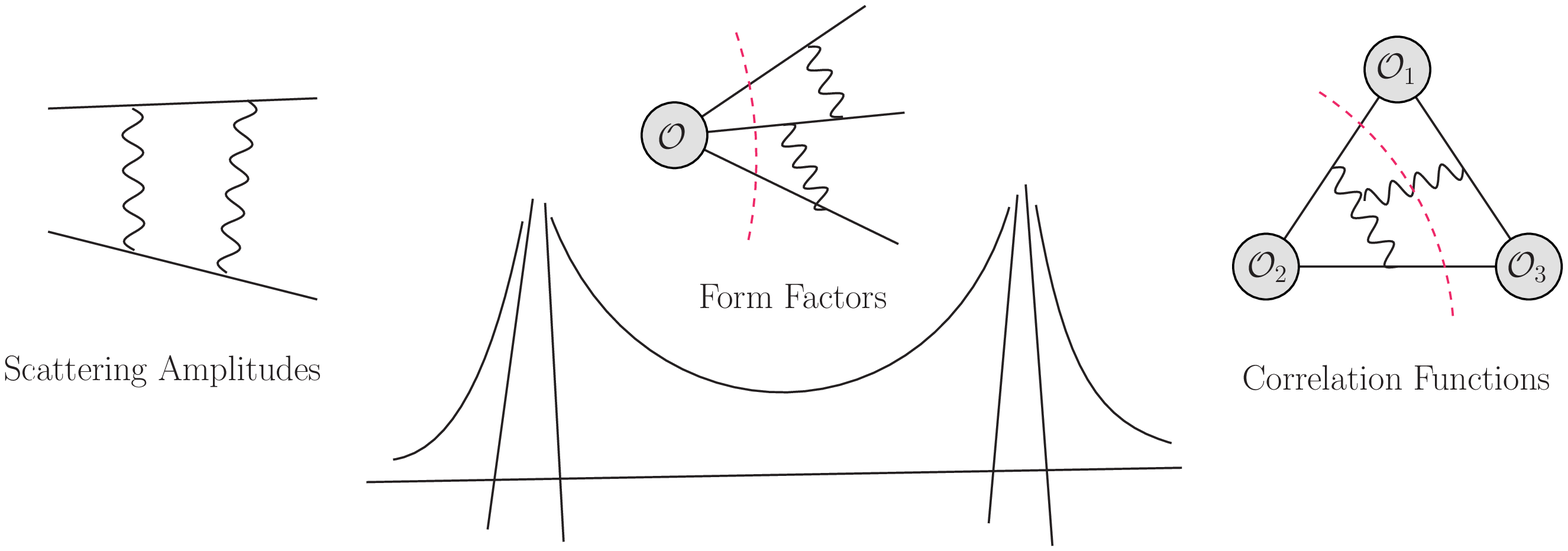}}
\caption{Form factors provide a bridge between amplitudes and correlation functions. By imposing on-shell unitarity cuts (indicated by the red dash lines), the amplitudes are building blocks in form factors, and so are form factors in correlation functions.} 
\label{fig:FFAmpCorr}
\end{figure}

Form factors are the matrix elements between on-shell asymptotic states and gauge invariant operators. 
The explicit definition of an $n$-point form factor can be given as
\begin{equation}
\label{eq:def-FF}
{\cal F}_{{\cal O},n} = \int d^D x \, e^{-i q \cdot x} \langle 1\cdots n|{\cal O}(x) | 0\rangle = (2\pi)^D\delta^{(D)}\Big(q-\sum_{i=1}^n p_i\Big)\langle 1\cdots n|{\cal O}(0) | 0\rangle ,
\end{equation}
where $p_i$ are the on-shell momenta of $n$ asymptotic particle states, and ${\cal O}$ is a local operator. By Fourier transformation, $q = \sum_i p_i$ is the off-shell momentum carried by the operator.
Therefore, form factors are partially on-shell and partially off-shell quantities, and they provide a natural bridge connecting the worlds of amplitudes and correlations functions,
as illustrated in Figure~\ref{fig:FFAmpCorr}.

%%%%%%%%%%%%%%%%%%%%%%%%%%%%%%%%%%
%%%%%%%%%%%%%%%%%%%%%%%%%%%%%%%%%%

In this review we will give an introduction to form factors in ${\cal N}=4$ super-Yang-Mills theory (SYM). 
${\cal N}=4$ SYM has been the primary model for the discovery and the developments of AdS/CFT correspondence \cite{Maldacena:1997re, Gubser:1998bc, Witten:1998qj}. 
It has also been a very important experimental ground for the modern amplitude developments.
The idea of unitarity cut method was first applied to compute one-loop amplitudes in ${\cal N}=4$ SYM in 1994 by Bern, Dixon, Durban and Kosower \cite{Bern:1994zx, Bern:1994cg}. 
In 2003 Witten's groundbreaking work provided a natural description of tree-level massless amplitudes using twistor space \cite{Witten:2003nn}. A key insight of this work that turned out to be very important later is the use of complex momentum and  three-point massless amplitudes. This soon lead to the discovery of several novel tree amplitude techniques: the BCFW recursion relation \cite{Britto:2004ap, Britto:2005fq}, the MHV rules \cite{Cachazo:2004kj} and the connected description \cite{Roiban:2004yf}. At the one-loop level, the generalized unitarity cut was developed based on three-point building blocks \cite{Britto:2004nc}. In 2005, based on the two- and three-loop 4-point computations in ${\cal N}=4$ SYM, Bern, Dixon and Smirnov proposed an ansatz of planar amplitudes to all orders \cite{Bern:2005iz}, which conjectures that a planar amplitude to all order can be given as the exponentiation of the one-loop correction. In 2007, Alday and Maldacena tested the BDS ansatz at strong coupling, where the computation is reduced to a geometric minimal surface problem using AdS/CFT correspondence \cite{Alday:2007hr}. This strong coupling picture, as well as the weak coupling observation \cite{Drummond:2006rz}, suggested the hidden dual conformal symmetry of planar amplitudes and the corresponding duality of planar amplitudes and null Wilson loops \cite{Drummond:2007aua, Brandhuber:2007yx, Drummond:2007au}. A review of this story can be found in \cite{Alday:2008yw}. 
Inspired by these developments and in particular Hodge's insight \cite{Hodges:2009hk},
hidden geometric structures such as Grassmannian and polytopes were discovered \cite{ArkaniHamed:2009dn, Mason:2009qx, ArkaniHamed:2009vw, ArkaniHamed:2012nw}. At the integrand level, the all-loop recursion relation was also developed \cite{ArkaniHamed:2010kv}. See \cite{Elvang:2013cua} for a review on these developments.
Another important progress is the discovery of a duality between color and kinematics by Bern, Carrasco and Johansson in 2008 \cite{Bern:2008qj, Bern:2010ue} which we will consider in more detail later.

Most of the above developments in amplitudes have been generalized to form factors. 
Although the Sudakov form factor in ${\cal N}=4$ SYM was first studied by van Neerven back in 1986 \cite{vanNeerven:1985ja},
it is only in recent years that the study of form factors in ${\cal N}=4$ SYM started to draw attention. This was first considered at strong coupling via AdS/CFT correspondence \cite{Alday:2007he, Maldacena:2010kp} and then at weak coupling \cite{Brandhuber:2010ad, Bork:2010wf, Brandhuber:2011tv, Bork:2011cj}, followed by many further studies. 
MHV structure of form factors and supersymmetric formalism were found in \cite{Brandhuber:2011tv, Penante:2014sza}. The duality between form factor and Wilson line, and the related dual conformal symmetry, were considered in \cite{Brandhuber:2010ad, Bianchi:2018rrj}. The strong coupling computation was generalized to full AdS$_5$ space using AdS/CFT correspondence \cite{Gao:2013dza}. Color-kinematics duality has been applied to compute form factors in \cite{Boels:2012ew, Yang:2016ear}. Grassmannian and polytopes pictures were studied in \cite{Frassek:2015rka, Bork:2016hst, Bork:2016xfn, Bork:2017qyh, Bork:2014eqa}. The twistor formalism was applied to form factors in \cite{Koster:2016ebi, Koster:2016loo, Chicherin:2016qsf, Koster:2016fna}. The connected description was developed in \cite{He:2016dol, Brandhuber:2016xue, He:2016jdg}. Recursion relation at the integrand level was studied in \cite{Bolshov:2018eos, Bianchi:2018peu}.
The unitarity computation of form factors of BPS operators was pursued in \cite{Brandhuber:2012vm, Brandhuber:2014ica}.
Form factors of non-protected operators and their application to the anomalous dimension problem in ${\cal N}=4$ SYM theory have been studied in \cite{Wilhelm:2014qua, Nandan:2014oga, Loebbert:2015ova, Brandhuber:2016fni, Loebbert:2016xkw, Caron-Huot:2016cwu}.
Form factors with multi-operator insertions have been studied in \cite{Engelund:2012re, Ahmed:2019upm, Ahmed:2019yjt} (see also a try at strong coupling in \cite{Gao:2013dza}).
Some other developments include \cite{Henn:2011by, Bork:2012tt, Johansson:2012zv, Huang:2016bmv, Ahmed:2016vgl}.
There have been several Ph.D theses devoted to the study of form factors using modern amplitudes techniques \cite{Engelund:2014laa, Gurdogan:2014gut, Wilhelm:2016izi, Penante:2016ycx, Koster:2017fvf, Meidinger:2018vlw}.  A brief review on form factors in ${\cal N}=4$ SYM can be found in \cite{Nandan:2018hqz}. 

It is beyond the scope of this review to cover all of above developments, 
and therefore we will focus on the aspects where the modern on-shell methods, in particular the unitarity methods, have been heavily used. 
One reason for this choice is that the on-shell formalism is based on general principles of quantum field theory. 
Therefore, although we consider the special ${\cal N}=4$ SYM, the ideas and techniques that we illustrate are expected to be applicable to general gauge theories. 
Having this in mind, we will restrict ourselves to the following aspects of form factors. The first class of form factors that we will consider are tree-level form factors (which are applicable to general massless gauge theories). Then we will consider two-point Sudakov form factors, which are important for computing infrared (IR) divergence quantities, such as the cusp and collinear anomalous dimensions. 
Finally, we consider loop form factors of generic non-BPS operators, from which we can extract ultraviolet (UV) anomalous dimensions.
Necessary background of amplitude techniques and pedagogical examples will be given. We expect the review to be understandable to a reader with a basic knowledge of quantum field theory.

\subsubsection*{Outline}

This review is structured as follows.

In Section~\ref{sec:preliminary}, 
we first give a brief review of the ${\cal N}$=4 SYM theory, then we take four-point tree amplitudes as examples to explain several important concepts and techniques: including traditional Feynman diagrams method (as a warm up), color-decomposition, color-kinematics duality, on-shell (super) spinor helicity formalism, as well as unitarity cut method.  

In Section~\ref{sec:treeFF}, we consider form factors at tree-level, which are the basic building blocks in the on-shell methods. The simple structure of tree results would be essential for the simplicity of loop quantities. As we will see, despite being partially off-shell, form factors preserve remarkable simple form as scattering amplitudes. Through minimal form factors, the spinor helicity formalism also provides a natural language for `on-shellize' generic local operators.

Section~\ref{sec:SudakovFF} is devoted to the two-point Sudakov form factor of a half-BPS operator, which contains important information of infrared divergences. As a quantity of single kinematic scale, it provides one of simplest examples to illustrate high loop computations. We explain the unitarity cut and color-kinematics duality method with explicit examples. 

In Section~\ref{sec:FFandAD}, we consider form factors with general non-protected operators. They contain UV divergences from which one can  compute anomalous dimensions of the operators. Examples in both SO(6) and SL(2) sectors are provided.

We conclude and give an outlook in Section~\ref{sec:conclusion}. 

Several technical details are given in appendices. The Feynman rules used in the main text are given in Appendix \ref{app:FeynRules}. The color factor computation is explained in Appendix \ref{app:coloralgebra}. We explain how to choose a basis of spinor products in Appendix \ref{app:spinorbasis}. The supersymmetric transformation is discussed in Appendix \ref{app:susy}. Finally, the integral convention and some formulas are given in Appendix \ref{app:integrals}.

%%%%%%%%%%%%%%%%%%%%%%%%%%%%%%%%%%%%
%%%%%%%%%%%%%%%%%%%%%%%%%%%%%%%%%%%%
%
\section{Preliminary: Four-Point Tree Amplitudes}
\label{sec:preliminary}

Many central concepts of modern amplitude studies can be understood by looking at the very simple four-point tree amplitudes. 
In this section, after a brief introduction of ${\cal N}=4$ SYM, we will use four-point amplitudes as major examples to explain several important ideas of amplitudes, including color-decomposition, color-kinematics duality, on-shell (super) spinor helicity formalism, and unitarity cut. 
This section is intended to give the necessary background of amplitudes, which will be used and generalized to study form factors in later sections.

%%%%%%%%%%%%%%%%%%%%%%%%%%%%%%
\subsection{${\cal N}$=4 super Yang-Mills theory}
\label{subsec:Neq4SYM}
${\cal N}$=4 SYM is the maximally supersymmetric massless gauge theory in four dimensions \cite{Brink:1976bc}. The field content consists of the same gauge boson $A_\mu$ as in quantum chromodynamics (QCD), plus extra particles: six real scalars $\Phi_I$ with $I = 1,..., 6$, and eight Weyl fermions $\Psi_{\alpha A}, \bar\Psi_{\dot\alpha}^A$ with $\alpha, \dot\alpha=1,2, A=1,2,3,4$. 
All fields are in the adjoint representation of the $SU(N_c)$ gauge group: 
\begin{equation}
A_\mu = A_\mu^a T^a \,, \quad \Phi_I = \Phi_I^a T^a \,, \quad \Psi = \Psi^a T^a \,,
\end{equation}
where $T^{a}$,  $a=1,\dots,N_{\text{c}}^2-1$, are the generators of $SU(N_c)$. The generators satisfy the commutation relation
\begin{equation}
[T^a, T^b] = i\sqrt{2} f^{abc}T^c  = \tilde f^{abc}T^c\,, \qquad \tilde f^{abc} = {\rm tr}(T^a T^b T^c) - {\rm tr}(T^a T^c T^b) \,,
\label{eq:fabc-def}
\end{equation} 
where $f^{abc}$ is the structure constant. 
We choose the normalization  ${\rm tr}(T^a T^b)= \delta^{ab}$.

The Lagrangian of ${\cal N}$=4 SYM can be written as
\begin{align}
L_{{\cal N}=4} = & {1\over g_{\rm YM}^2} {\rm Tr} \bigg[ {1\over4}F_{\mu\nu}^2 + {1\over2} (D_\mu\Phi_I)^2 - {1\over4} [\Phi_I, \Phi_J]^2 + \textrm{fermionic part} \bigg] \,,
\end{align}
where the covariant derivative is 
\begin{equation}
D_\mu\, \cdot = \partial_\mu \, \cdot - i g_{\rm YM} [A_\mu, \, \cdot] \,.
\end{equation}
The details of the fermionic part will not be needed in this review. From the Lagrangian one can derive Feynman rules, and we collect the Feynman rules for gluons and scalars in Appendix~\ref{app:FeynRules}.

${\cal N}=4$ SYM is a special theory in the sense that it enjoys many symmetries.
First, the theory is maximally supersymmetric: the bosonic fields, 2 gluons plus 6 scalars, have eight physical degrees of freedom which match the number of fermions. This corresponds to a global R-symmetry $SU(4)_R \simeq SO(6)_R$. Furthermore, the $\beta$ function is zero to all order, thus the theory is a conformal field theory (CFT) \cite{Sohnius:1981sn, Green:1982sw, Mandelstam:1982cb, Brink:1982wv}.  Supersymmetry and conformal symmetry together form a larger global symmetry group $PSU(2,2|4)$. Moreover, in the `t Hooft large $N_c$ limit \cite{tHooft:1973jz}, the planar ${\cal N}$=4 SYM contains infinitely many hidden symmetries which render the theory integrable, see an extensive review in \cite{Beisert:2010jr}.

Since ${\cal N}=4$ SYM is a CFT, a commonly raised question is: are scattering amplitudes well defined in this theory?  This seems indeed puzzling since the asymptotic states are not physically well-defined in CFT. This problem may be solved from several viewpoints. 
First, there is certainly no problem to formally define a perturbative S-matrix from a given Lagrangian. In particular, the tree-level amplitudes take universal form that are not sensitive to which theory one is considering, for example, the gluon tree amplitudes are identical in ${\cal N}=4$ SYM and QCD. 
Second, at loop level, the loop integrals contain divergences. This requires to introduce a regulator, such as changing dimension to be $D=4-2\epsilon$ in dimensional regularization, where the conformal symmetry is broken. 
Last but not least, amplitudes are \emph{not yet} the final physical observables. In this respect, one should note that even in QCD, there are also no asymptotic free gluons or quarks because of confinement, but one can still consider QCD amplitudes by assuming there are free on-shell partons. (In QCD, the factorization is the underlying mechanism that justifies this assumption, see e.g. \cite{Collins:2011zzd}.)
Similar point of view can be taken for ${\cal N}=4$ SYM.  

%%%%%%%%%%%%%%%%%%%%%%%%%%%%%%
\subsection{Four-point amplitudes}

Now we start to review amplitudes techniques using four-point tree amplitudes as concrete examples.
As a warm-up, we first review the computation of standard Feynman diagrams. 

%%%%%%%%%%%%%%%%%%%%%%%%%%%%%%
\subsubsection*{The four-gluon amplitude}

%%%%%%%%%%%%%%%%%%%%%%%%%%%%%%%%%%
\begin{figure}[t]
\centerline{\includegraphics[height=2.5cm]{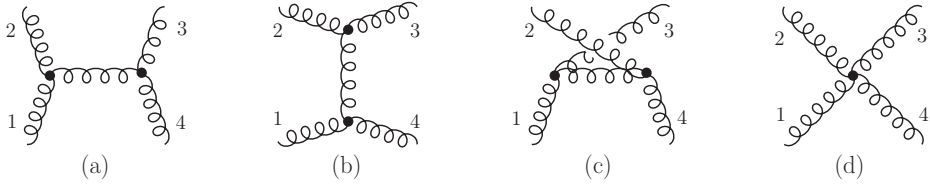} } 
\caption{Feynman diagrams of the four-gluon tree amplitude.} 
\label{fig:FeynDiagA4gluon}
\end{figure}
%%%%%%%%%%%%%%%%%%%%%%%%%%%%%%%%%%

There are four Feynman diagrams for the four-gluon tree amplitude as given in Figure~\ref{fig:FeynDiagA4gluon}. 
Using the Feynman rules given in Appendix \ref{app:FeynRules}, one obtains the amplitude as a function of color factors and Lorentz products of momenta and polarization vectors:\footnote{We will in general neglect the gauge coupling $g_{\rm YM}$ which can be easily recovered.} 
\begin{align}
{\hat A}^{(0)}_4 (\{\varepsilon_i, p_i, a_i\}) = & - i f^{a_1 a_2 b} f^{a_3 a_4 b} \Big\{  {\eta_{\mu\nu}\over s_{12}}  \big[ (\varepsilon_1\cdot\varepsilon_2)(p_1^\mu - p_2^\mu) + 2 (\varepsilon_1 \cdot p_2) \varepsilon_2^\mu - 2 (\varepsilon_2\cdot p_1)\varepsilon_1^\mu \big] \nonumber\\
& \hskip 3.6cm \times \big[  (\varepsilon_3\cdot\varepsilon_4)(p_{3}^{\nu} - p_{4}^{\nu}) + 2 (\varepsilon_3 \cdot p_4) \varepsilon_{4}^{\nu} - 2 (\varepsilon_4\cdot p_3)\varepsilon_{3}^{\nu} \big] \nonumber\\
& \hskip 2.9cm+ (\varepsilon_1 \cdot \varepsilon_3)(\varepsilon_2 \cdot \varepsilon_4)-(\varepsilon_1 \cdot \varepsilon_4)(\varepsilon_2 \cdot \varepsilon_3)  \Big\} \nonumber\\
& \ + (1\leftrightarrow3) + (2\leftrightarrow3)  \,.
\label{eq:A4fullcolor}
\end{align}

Expanding the structure constants $f^{abc}$ in terms of $T^a$'s as \eqref{eq:fabc-def}, and contracting color indices using the completeness relation (see Appendix \ref{app:coloralgebra} for more discussion)
\begin{equation}
\sum_{a=1}^{N_c^2-1} (T^a)_i^{~j} (T^a)_k^{~l} = \delta_i^{~l} \delta_k^{~j} - {1\over N_c} \delta_i^{~j} \delta_k^{~l} \,,
\label{eq:colorcontra}
\end{equation}
the four-gluon amplitude can be reorganized as a sum of six terms:
\begin{align}
{\hat A}^{(0)}_4 =& {\rm tr}(T^{a_1} T^{a_2} T^{a_3}  T^{a_4}) A^{(0)}_4(1,2,3,4) +  {\rm tr}(T^{a_1} T^{a_3} T^{a_2}  T^{a_4}) A^{(0)}_4(1,3,2,4) \nonumber\\
& + {\rm tr}(T^{a_2} T^{a_1} T^{a_3}  T^{a_4}) A^{(0)}_4(2,1,3,4) +  {\rm tr}(T^{a_2} T^{a_3} T^{a_1}  T^{a_4}) A^{(0)}_4(2,3,1,4) \nonumber\\
& + {\rm tr}(T^{a_3} T^{a_1} T^{a_2}  T^{a_4}) A^{(0)}_4(3,1,2,4) +  {\rm tr}(T^{a_3} T^{a_2} T^{a_1}  T^{a_4}) A^{(0)}_4(3,2,1,4) \nonumber\\
=& \sum_{\sigma\in S_3} {\rm tr}(T^{a_{\sigma(1)}} \cdots T^{a_{\sigma(3)}}  T^{a_4}) A^{(0)}_4({\sigma(1)},{\sigma(2)},{\sigma(3)},4) \,,
\label{eq:hatA4decomposition}
\end{align}
where each term contains a single-trace color factor and a component $A_4$ which depends only on kinematics. We will call $A_4$ the color-ordered amplitudes. One can check that $A_4(1,2,3,4)$ in the first term in \eqref{eq:hatA4decomposition} can be computed using three planar Feynman diagrams, (a), (b), (d) given in Figure~\ref{fig:FeynDiagA4gluon}, with color-stripped Feynman rules given in Appendix \ref{app:FeynRules}.

Similarly, an $n$-gluon amplitude can be decomposed as
\begin{equation}
{\hat A}_{n}(\{a_i, p_i, \eta_i\}) = \sum_{\sigma\in S_n/Z_n} {\rm Tr}(T^{a_{\sigma(1)}} \cdots T^{a_{\sigma(n)}}) A_{n}(\{p_{\sigma(i)}, \eta_{\sigma(i)}\}) + \text{multi-trace terms} \,.
\end{equation}
The first sum contains only color single-trace contributions, while the multi-trace terms contribute only starting from one and higher loop orders.
We denote ${\hat A}_n$ for amplitudes with full color dependence, and $A_n$ for color-ordered amplitudes.

\begin{figure}[t!]
\centering
\includegraphics[height=2.4cm]{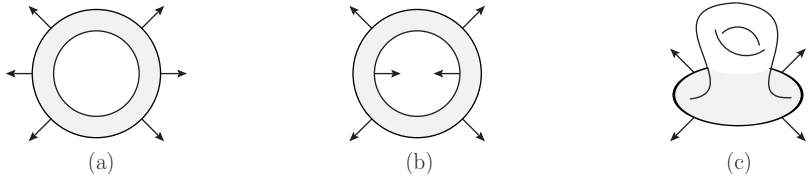}
\caption{Examples of planar and non-planar topologies: (a) is a single-trace one-loop planar diagram, (b) is a double-trace one-loop diagram, and (c) is a single trace but non-planar diagram. Both (b) and (c) are suppressed in the large $N_c$ limit.} 
\label{fig:stringdiag}
\end{figure}

The color decomposition has a natural interpretation of planarity in the `t Hooft  large $N_c$ limit \cite{tHooft:1973jz}, in which one takes $N_c$ to be infinitely large while keeping $\lambda = g_{\rm YM}^2 N_c$ fixed. 
In this limit, only the single-trace terms survive. Therefore, the corresponding components, the color-ordered amplitudes with leading $N_c$ factor, will be also called planar amplitudes. They can be computed using planar Feynman diagrams with color-stripped Feynman rules. 
We illustrate some examples of planar and non-planar topologies in Figure~\ref{fig:stringdiag}.

In Section~\ref{sec:CKduality}, we will find that amplitudes ${\hat A}$ of full color dependence can have some hidden structure (a duality between color factors and kinematics factors), which is not visible in the planar limit. Therefore, it is important to keep in mind that: while planar color decomposition is an important way of simplification, considering the full color dependence may also have significant advantages. 

%%%%%%%%%%%%%%%%%%%%%%%%%%%%%%%%%%
%%%%%%%%%%%%%%%%%%%%%%%%%%%%%%%%%%

%%%%%%%%%%%%%%%%%%%%%%%%%%%%%%
\subsubsection*{The four-scalar amplitudes}

As a further example in ${\cal N}=4$ SYM, we consider the tree amplitudes with four external scalars.
They will be used as building blocks to compute the anomalous dimensions in the SO(6) sector in Section~\ref{sec:FFandAD}.
Since the color structure is same as that of the gluon amplitudes, we only consider the planar color-ordered amplitudes below.

%%%%%%%%%%%%%%%%%%%%%%%%%%%%%%%%%%
\begin{figure}[t]
\centering
\includegraphics[height=2.5cm]{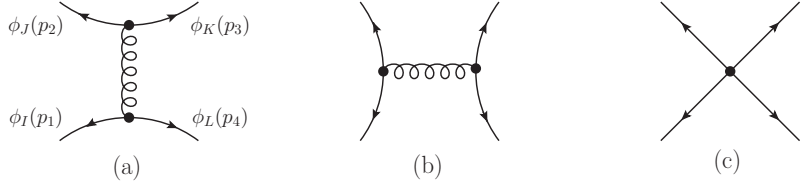} 
\caption{Planar Feynman diagrams for the four-scalar color-stripped amplitudes.} 
\label{fig:FeynDiagA4scalar}
\end{figure}
%%%%%%%%%%%%%%%%%%%%%%%%%%%%%%%%%%

\begin{figure}[t]
\centerline{\includegraphics[height=2.2cm]{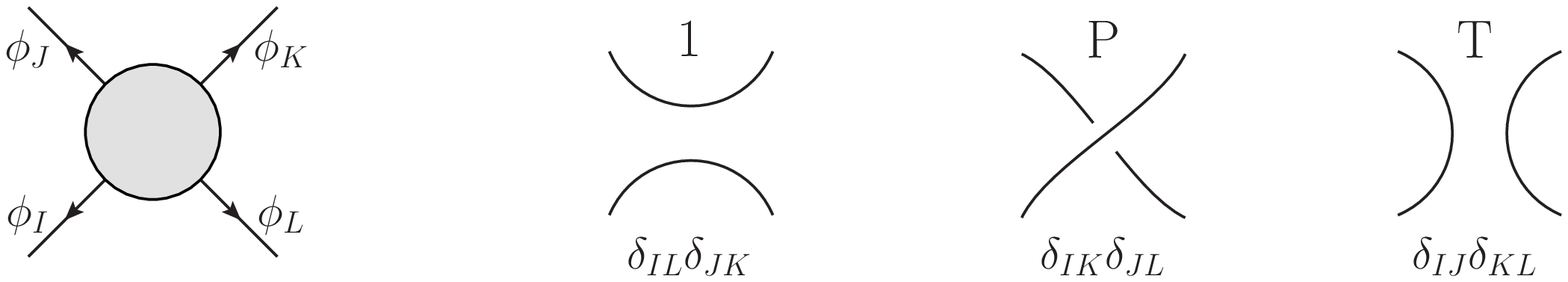} } 
\caption{R-charge flow structure for the four-scalar amplitudes.} 
\label{fig:A4scalar_IPT}
\end{figure}

The planar four-scalar amplitudes have three Feynman diagram contributions, as shown in Figure~\ref{fig:FeynDiagA4scalar}:
\begin{equation}
A(1_{\phi_I}, 2_{\phi_J}, 3_{\phi_K}, 4_{\phi_L}) = {\rm Diag.\,(a)} + {\rm Diag.\,(b)} + {\rm Diag.\,(c)} \,.
\end{equation}
The first two diagrams are related by symmetry. Using Feynman rules in Appendix \ref{app:FeynRules}, one has
\begin{align}
{\rm Diag.\,(a)} & =  -{i \over (p_1 + p_4)^2} \Big({i \over \sqrt{2}}\Big)^2 (p_4 - p_1) \cdot (p_2 - p_3) \delta_{IL}\delta_{JK}  = -i \bigg( {1 \over 2} + {s_{12} \over s_{1 4}}  \bigg)  \delta_{IL}\delta_{JK} \,, \\
{\rm Diag.\,(b)} & =  - i \bigg( {1 \over 2} + {s_{14} \over s_{12}} \bigg)  \delta_{IJ}\delta_{KL} \,, \qquad s_{ij} := (p_i+p_j)^2 \,.
\end{align}
The third diagram is simply a four-scalar self-interaction vertex which gives
\begin{equation}
{\rm Diag.\,(c)} = -{i \over 2}  \delta_{IL}\delta_{JK} + i \delta_{IK}\delta_{JL}  - {i\over 2} \delta_{IJ}\delta_{KL} \,.
\end{equation}
In total, the planar four-scalar amplitudes are
\begin{equation}
A(1_{\phi_I}, 2_{\phi_J}, 3_{\phi_K}, 4_{\phi_L}) = i \left[ \left( -1 -{s_{12} \over s_{1 4} } \right) \mathds{1} + \mathds{P} + \left(-1 - {s_{1 4} \over s_{12}} \right) \mathds{T} \right] \,,
\label{eq:A4scalar1}
\end{equation}
where we introduce
\begin{equation}
\mathds{1} := \delta_{IL}\delta_{JK} \,, \quad \mathds{P} := \delta_{IK}\delta_{JL} \,, \quad \mathds{T} := \delta_{IJ}\delta_{KL} \,. 
\end{equation}
This implies that the four-scalar amplitudes have three types of R-change flow structures,  which are illustrated in  Figure~\ref{fig:A4scalar_IPT}. 
They have a nice physical picture which will be discussed in Section \ref{sec:SO6AD}.

%%%%%%%%%%%%%%%%%%%%%%%%%%%%%%%%%%
%%%%%%%%%%%%%%%%%%%%%%%%%%%%%%%%%%

%%%%%%%%%%%%%%%%%%%%%%%%%%%%%%
%%%%%%%%%%%%%%%%%%%%%%%%%%%%%%
\subsection{Color-kinematics duality}
\label{sec:CKduality}

A surprising duality between color and kinematics in amplitudes was discovered by Bern, Carrasco and Johansson \cite{Bern:2008qj, Bern:2010ue}.
The duality conjectures that there exists a cubic graph representation of amplitudes in which the kinematic numerators satisfy equations in one-to-one correspondence with Jacobi relations of the corresponding color factors. This indicates a deep connection between the kinematic and color structures in gauge theories.

%%%%%%%%%%%%%%%%%%%%%%%%%%%%%%%%%%
\begin{figure}[h]
\centerline{\includegraphics[height=2cm]{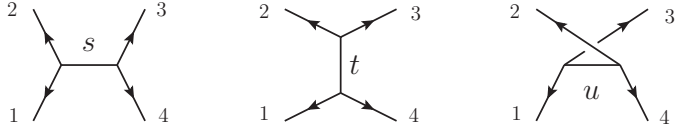} } 
\caption{Trivalent graphs of four-point tree amplitudes.} 
\label{fig:treeA4}
\end{figure}
%%%%%%%%%%%%%%%%%%%%%%%%%%%%%%%%%%

A basic and important example to understand the color-kinematics duality is the four-point tree amplitudes. 
The starting point is to express four-point amplitudes as a sum of three terms, 
\begin{align}
\label{eq:BCJ-4pt-tree}
\hat A^{\rm tree}_4(1, 2, 3, 4) \, = \, {c_s \, n_s \over s} + {c_t\, n_t \over t} + {c_u \, n_u \over u} \,,
\end{align}
which correspond to three trivalent topologies in Figure~\ref{fig:treeA4}.
The $c_i$ are color factors defined by the product of structure constants $\tilde f^{abc}$ associated to each trivalent vertex:
\begin{align} 
c_s =\tilde f^{a_1 a_2 b}\tilde f^{b a_3 a_4}, \quad 
c_t =\tilde f^{a_2 a_3 b}\tilde f^{b a_4 a_1}, \quad 
c_u =\tilde f^{a_1 a_3 b}\tilde f^{b a_2 a_4} \,,
\end{align}
which satisfy the Jacobi relation:
\begin{equation} 
c_s = c_t + c_u \,.
\end{equation}
The numerators $n_i$ are kinematic factors depending on the polarization and momentum invariants.
Since the color factors and the propagators are simply determined by the topologies, the genuine non-trivial information is contained in the kinematic factors $n_i$.
The color-kinematics duality requires that the numerators satisfy the same Jacobi relation of color factors:
\begin{align} 
c_s = c_t + c_u  \ \  \Rightarrow \ \  n_s = n_t + n_u \,,
\label{eq:jacobi-rel}
\end{align}
which we will refer as dual Jacobi relation or kinematic Jacobi relation.

One can check if the dual Jacobi relation is correct using the Feynman diagram results in last subsection. 
From the four-gluon amplitude result  \eqref{eq:A4fullcolor} and collecting terms according the color factors, one finds
\begin{align}
n_s^{4\textrm{-gluon}} = & {i\over2} \eta_{\mu\nu}  \big[ (\varepsilon_1\cdot\varepsilon_2)(p_1^\mu - p_2^\mu) + 2 (\varepsilon_1 \cdot p_2) \varepsilon_2^\mu - 2 (\varepsilon_2\cdot p_1)\varepsilon_1^\mu \big] \nonumber\\
& \qquad  \times \big[  (\varepsilon_3\cdot\varepsilon_4)(p_{3}^{\nu} - p_{4}^{\nu}) + 2 (\varepsilon_3 \cdot p_4) \varepsilon_{4}^{\nu} - 2 (\varepsilon_4\cdot p_3)\varepsilon_{3}^{\nu} \big] \nonumber\\
& + s_{12} (\varepsilon_1 \cdot \varepsilon_3)(\varepsilon_2 \cdot \varepsilon_4)-(\varepsilon_1 \cdot \varepsilon_4)(\varepsilon_2 \cdot \varepsilon_3)  \,,
\end{align}
and the $t$ and $u$-channel numerators are related by symmetry as
\begin{align}
n_t = n_s |_{1\leftrightarrow 3} \,, \qquad n_u = n_s |_{2\leftrightarrow 3} \,.
\label{eq:ntnsbysym}
\end{align}
It is then straightforward to check the kinematic numerators indeed satisfy the dual Jacobi relation \eqref{eq:jacobi-rel}. Note that one needs to use the on-shell conditions $p_i^2 = \varepsilon_i \cdot p_i=0$.

Similarly, for four-scalar amplitudes, one has
\begin{align}
n_s^{4\textrm{-scalar}} = & {i\over2} \big[ (p_1 - p_2) \cdot (p_3 - p_4) \delta_{I_1 I_2} \delta_{I_3 I_4} + s_{12} (\delta_{I_1 I_3} \delta_{I_2 I_4} - \delta_{I_1 I_4} \delta_{I_2 I_3}) \big]   \,,
\end{align}
and $n_{t}, n_u$ are similarly given by \eqref{eq:ntnsbysym}.  They also satisfy the dual Jacobi relation.

Although the duality is satisfied directly for four-point amplitudes, this is not obvious to be true for high-point or high-loop cases. For general tree-level amplitudes, the existence of such a representation has been proved based on monodromy relations in string amplitudes \cite{BjerrumBohr:2009rd, Stieberger:2009hq} or directly in field theory using the BCFW recursion relation \cite{Feng:2010my}. At the loop level, the general existence of this duality is still a conjecture and  relies on a case-by-case proof. 
In self-dual YM/gravity, the dual Jacobi relation can be nicely interpretated as the algebra of area-preserving diffeomorphism \cite{Monteiro:2011pc}. 
An important bonus of the duality is its power in computing gravity amplitudes \cite{Bern:2010ue, Bern:2010yg}, which is dubbed the \emph{double copy} construction and is closed related to the KLT relation \cite{Kawai:1985xq} and CHY formalism \cite{Cachazo:2013hca, Cachazo:2013iea}.
We will not discuss this aspect in this review. We suggest interested reader to \cite{Bern:2019prr} for an introduction and extensive review of the subject.
We will discuss the loop construction of Sudakov form factors using color-kinematics duality in Section~\ref{sec:SudakovFF}.

%%%%%%%%%%%%%%%%%%%%%%%%%%%%%%
%%%%%%%%%%%%%%%%%%%%%%%%%%%%%%
\subsection{Spinor helicity formalism}
\label{sec:spinorhelicity}
The central idea of on-shell method is to compute the amplitudes with simpler on-shell quantities. Starting with simplest tree-level building blocks, one can construct higher-point tree amplitudes via recursion relations based on the analytic and factorization properties of amplitudes \cite{Britto:2004ap, Britto:2005fq}. Using (generalized) unitarity cut, loop amplitudes can be further constructed using tree-level building blocks \cite{Bern:1994zx, Bern:1994cg, Britto:2004nc}. 
These ideas can be best realized using the spinor helicity formalism \cite{Xu:1986xb, DeCausmaecker:1981jtq, Berends:1981rb, Kleiss:1985yh}. We provide a brief introduction of this formalism below, and reader can find more extensive discussion in e.g.~\cite{Dixon:1996wi, Elvang:2013cua, Henn:2014yza}.

The basic idea is to express four-momentum as a product of two spinors. For the massless momentum $p_j$ (of $j$th particle), one has
\begin{equation}
p_j^{\alpha\dot\alpha} = p_j^\mu \sigma_\mu^{\alpha\dot\alpha} = \lambda_j^\alpha \tilde\lambda_j^{\dot\alpha},
\end{equation}
where $\sigma_\mu = ({\bf 1}, \sigma_i)$ with $\sigma_i$ the Pauli matrices. The new variables $\lambda^\alpha, \tilde\lambda^{\dot\alpha}$ are 2-component Weyl spinors, with $\alpha,\dot\alpha=1,2$. The Lorentz product of four-momenta is given in terms of spinor contraction as
\begin{equation}
2p_i\cdot p_j = \langle i\,j \rangle [j\,i], \qquad \langle i\,j\rangle = \epsilon_{\alpha\beta} \lambda_i^\alpha\lambda_j^\beta, \quad [ i \, j] = \epsilon^{\dot\alpha\dot\beta}  \tilde\lambda_{i,\dot\alpha} \tilde\lambda_{j,{\dot\beta}} \,,
\end{equation}
where $\epsilon_{\alpha\beta}, \epsilon^{\dot\alpha\dot\beta}$ are antisymmetric tensors and can be used to bring down or up the spinor indices. The spinor products satisfy the Schouten identity:
\begin{equation}
\langle i\,j\rangle \langle k\,l\rangle + \langle k\,i\rangle \langle j\,l\rangle + \langle j\,k\rangle \langle i\,l\rangle  =0 \,,
\label{eq:schouten}
\end{equation}
with a similar one for $\tilde\lambda$ by changing $\langle \,\rangle \rightarrow [\,]$.
Because of such  non-linear relations, expressions given in terms of spinor products are not unique and it is in general hard to simplify a result systematically.
In this respect, it is convenient to choose an independent basis of spinor products. This is explained in Appendix \ref{app:spinorbasis}.

There are several advantages by introducing the spinor representation. 
First of all, the polarization vectors of gluons can be expressed in the bi-spinor representation as
\begin{equation}
\varepsilon^{(+)}_{i,\alpha\dot\alpha} = \sqrt{2} \, {\xi_{i,\alpha} \tilde\lambda_{i,\dot\alpha} \over \langle \xi_i\lambda_i \rangle} \,, \qquad 
\varepsilon^{(-)}_{i,\alpha\dot\alpha} = \sqrt{2} \, {\lambda_{i,\alpha} \tilde\xi_{i,\dot\alpha} \over [\lambda_i\xi_i]} \,,
\label{eq:polarization}
\end{equation}
where $\xi, \tilde\xi$ are arbitrary reference spinors.

Second, it allows a simple representation of three-point amplitudes.
This is not trivial since the physical three-gluon amplitude is zero. To obtain a non-trivial expression, one needs to go to complex momentum space which is easy to realize using spinor variables \cite{Witten:2003nn}:
\begin{align}
A_3(1^-,2^-,3^+) = i{\langle 12 \rangle^3 \over \langle 23 \rangle \langle 31 \rangle} \,, \qquad A_3(1^+,2^+,3^-) = -i {[ 12 ]^3 \over [ 23 ] [ 31 ]} \,.
\label{eq:A3spinorform}
\end{align}
The above three-gluon amplitudes can be fixed by dimensional analysis and little group transformation, see e.g.~\cite{Arkani-Hamed:2017jhn}.
It is also possible to obtain this form using Feynman diagram expression
\begin{equation}
A_3 = i \sqrt{2} \big[ (\varepsilon_1 \cdot p_2) (\varepsilon_2 \cdot \varepsilon_3) + (\varepsilon_2\cdot p_3) (\varepsilon_3 \cdot \varepsilon_1) + (\varepsilon_3\cdot p_1) (\varepsilon_1\cdot\varepsilon_2) \big] \,,
\end{equation}
and the spinor form of polarization vectors \eqref{eq:polarization}.\footnote{It may be interesting to note that one could easily find the form of \eqref{eq:A3spinorform} by computing three-gluon form factor of ${\rm tr}(F^2)$ operator and then taking the $q^2\rightarrow0$ limit.}

Third, the spinor helicity formalism allows us to decompose amplitudes according to their helicity configurations, which is usually called {\it helicity amplitudes}. The simplest class of helicity amplitudes are the maximally helicity violating (MHV) gluon amplitudes which have only two negative helicities. They take the remarkably simple Parke-Taylor form \cite{Parke:1986gb}:
\begin{equation}
A_n(1^+ \ldots, i^-, \ldots j^-, \ldots, n^+)   = i { \langle ij\rangle^4  \over \langle 12\rangle \ldots \langle n1\rangle} \,.
\label{eq:MHVamplitude}
\end{equation}
Such kind of simplicity is not obvious at all using Feynman diagram expressions. 
With MHV amplitudes as building blocks, one can systematically compute other non-MHV helicity amplitudes using e.g.~MHV rules \cite{Cachazo:2004kj}. 

Since we consider supersymmetric ${\cal N}=4$ SYM, we also introduce super spinor helicity formalism. The on-shell supermomentum $Q_j^{\alpha A}$ of the $j$th external partcle can be defined as
\begin{equation}
Q_j^{\alpha A} = \lambda_j^\alpha \eta_j^A,
\end{equation}
where $\eta^A_j$'s are Grassmann variables and the superscripts $A = 1, \dots, 4$ denote the $SU(4)_R$ indices. 
The on-shell states can be described by the ${\cal N}=4$ on-shell superfield \cite{Nair:1988bq}:
\begin{equation}
W(p,\eta)=g_+(p) +  \eta^A \, \bar\psi_A(p) + {\eta^A\eta^B \over 2!} \, \phi_{AB}(p) + { \epsilon_{ABCD} \eta^A\eta^B\eta^C \over 3!} \, \psi^D(p) + \eta^1\eta^2\eta^3\eta^4 \, g_-(p) \,.
\label{eq:onshellN=4superspace}
\end{equation}
In \eqref{eq:onshellN=4superspace}, the six real scalars $\phi_I$ are changed from the $SO(6)_R$ to the $SU(4)_R$ representation as $\phi_{AB} = \Sigma^I_{AB} \phi_I$, which satisfy
\begin{equation}
\phi_{AB} = - \phi_{BA} \,, \qquad \phi^{AB} = {1\over2} \epsilon^{ABCD} \phi_{CD} \,.
\end{equation}
The on-shell superfields fulfill a representation of the ${\cal N}=4$ superalgebra, which is discussed in Appendix \ref{app:susy}.

With the on-shell superfield representation, amplitudes of different external particles can be written in a very compact form. The super MHV amplitudes can be given as
\begin{equation}
\label{eq:AsuperMHV}
{\cal A}_n^{\rm MHV} = i {\delta^{(4)}(\sum_i \lambda_i\tilde\lambda_i) \delta^{(8)}(\sum_i \lambda_i \eta_i) \over \langle 12\rangle \langle 23\rangle \ldots  \langle n 1 \rangle} \,,
\end{equation}
where the two delta functions in the numerator correspond to momentum conservation and super-momentum conservation, respectively. 
To get amplitudes of specific external particles, we can extract corresponding components in the $\eta$ expansion. For example, using the expansion formula:
\begin{equation}
\delta^{(8)}(\sum_i \lambda_i^\alpha \eta_i^A) = \prod_{A=1}^4 \prod_{\alpha=1}^2 (\sum_i \lambda_i^\alpha \eta_i^A) = \prod_{A=1}^4 (\sum_{i<j} \langle i j \rangle \eta_i^A \eta_j^A) \,,
\end{equation}
and taking the components of $\eta_1^1 \eta_1^2 \eta_1^3 \eta_1^4 \eta_2^1 \eta_2^2 \eta_2^3 \eta_2^4$ and $\eta_1^1 \eta_1^2 \eta_2^1 \eta_2^2 \eta_3^3 \eta_3^4 \eta_4^3 \eta_4^4$ respectively, we obtain the four-gluon and four-scalar amplitudes:
\begin{align}
& A_4(1^-, 2^-, 3^+, 4^+)   = i { \langle 12\rangle^4  \over \langle 12\rangle \langle 23\rangle \langle 34\rangle \langle 41\rangle}  \,, \\
& A_4(1_{\phi_{12}}, 2_{\phi_{12}}, 3_{\phi_{34}}, 4_{\phi_{34}})   = i { \langle 12\rangle^2 \langle 34\rangle^2  \over \langle 12\rangle \langle 23\rangle \langle 34\rangle \langle 41\rangle}  \,.
\end{align}
We use ${\cal A}$ (${\cal F}$) to represent supersymmetric planar amplitudes (form factors) and $A$ ($F$) for non-supersymmetric component planar amplitudes (form factors).

As an analogy of \eqref{eq:A4scalar1}, we can write the four-scalar amplitudes in the $SU(4)_R$ representation as:
\begin{equation}
A_4(1_{\phi_{AB}}, 2_{\phi_{CD}}, 3_{\phi_{C'D'}}, 4_{\phi_{A'B'}})  = i \left[ {\langle 13\rangle \langle 24\rangle \over \langle 14\rangle \langle 23\rangle}  \mathds{1} + \mathds{P} + {\langle 13\rangle \langle 24\rangle \over \langle 12\rangle \langle 34\rangle} \mathds{T} \right]  \,,
\end{equation}
where
\begin{equation}
\mathds{1} := \epsilon_{ABA'B'} \epsilon_{CDC'D'} \,, \quad \mathds{P} := \epsilon_{ABC'D'} \epsilon_{CDA'B'} \,, \quad \mathds{T} := \epsilon_{ABCD} \epsilon_{A'B'C'D'} \,.
\end{equation}

On-shell spinor helicity formalism can be also applied to form factors in which the new building blocks are the minimal form factors, which will be introduced in Section~\ref{sec:mimiFF}.

%%%%%%%%%%%%%%%%%%
\subsection{On-shell unitarity cut}

S-matrix in QFT is an analytic function of kinematics and can be determined by the {\it singularities}, i.e., poles and branch cuts. All such singularities have physical origins and can be understood through the unitarity property of the S-matrix: poles are related to particle-fusion to a single particle, and branch cuts are related to the thresholds of multi-particle generation. 
It has been hoped that via analytic properties based on general physical principles, amplitudes may be determined completely, which is known as the S-matrix bootstrap program, see e.g.~\cite{Gribov:2003nw}.
Back in 1960s (the pre-QCD era), this strategy was applied with the hope of understanding strong interactions. 

The modern unitarity method works at perturbative level \cite{Bern:1994zx, Bern:1994cg, Britto:2004nc}. It provides a powerful practical tool of constructing perturbative amplitudes through performing {\it cut}, which corresponds to set the momentum of an internal propagator to be on-shell:
\begin{equation}
{i \over l^2} \  \stackrel{\rm cut}{\longrightarrow} \  2\pi \delta_+(l^2) \,.
\label{eq:cutdef}
\end{equation}
Under cuts,  an amplitude can factorize as products of simpler amplitudes, i.e.~those of lower loops or with fewer external legs.

%%%%%%%%%%%%%%%%%%%%%%%%%%%%%%%%%%
\begin{figure}[h]
\centerline{\includegraphics[height=2.cm]{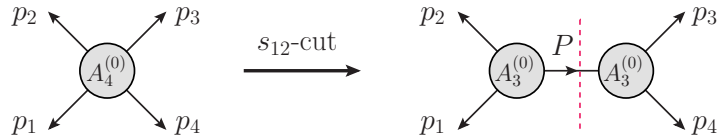} } 
\caption{The $s$-channel unitarity cut of four-point tree amplitudes.} 
\label{fig:A4s12cut}
\end{figure}
%%%%%%%%%%%%%%%%%%%%%%%%%%%%%%%%%%

A simple example to understand unitarity cut is again the four-point tree amplitudes. As show in Figure~\ref{fig:A4s12cut}, one can perform a $s_{12}$-cut, such that the four-point amplitude factorizes into two three-point amplitudes.
One can check the following relation using Feynman diagram expression for planar gluon amplitudes $A_4$ and $A_3$:\footnote{The factor $1/2$ is added since for fixed momenta, only one helicity configuration is nonzero, which can be also understood from the 3-point helicity amplitudes.}
\begin{equation}
\lim_{s_{12} \rightarrow0} s_{12} A_4^{(0)}(1,2,3,4) = {1\over2} \sum_\textrm{helicity of $\varepsilon(P)$} A_3^{(0)}(1,2,P) A_3^{(0)}(-P,3,4)  \,,
\end{equation} 
where the helicity sum for the cut leg $P$ can be performed using:
\begin{equation}
\sum_{\rm helicities}\varepsilon^{\mu}(P) \,\varepsilon^{\nu}(P)=\eta^{\mu\nu}
-\frac{q^{\mu}P^{\nu}+q^{\nu}P^{\mu}}{q\cdot P} \,.
\label{eq:helicity-contraction-rule}
\end{equation}
Similar expression holds if one uses helicity amplitudes, where one sums all possible $\pm$ helicities. 

If we use super amplitudes in ${\cal N}=4$ SYM, the sum of all possible internal states can be done via the integration of Grassmann variables $\eta_P$:
\begin{equation}
\qquad \sum_\textrm{helicity $\varepsilon(P)$} \ \rightarrow \ \int d^4 \eta_P \,, \qquad \textrm{where }\int d^4 \eta_P \, \eta_P^1 \eta_P^2 \eta_P^3 \eta_P^4 = 1 \,.
\label{eq:etaIntegral}
\end{equation} 

For loop amplitudes, one can perform similar cuts for the loop integrand, which has no much difference comparing to the above example, except that in general multiple cuts are required.
Using tree building blocks, one can reconstruct the loop amplitudes in a much more efficient way.
We will consider explicit examples of loop form factors in Section~\ref{sec:SudakovFF} and \ref{sec:FFandAD}.

%%%%%%%%%%%%%%%%%%%%%%%%%%%%%%%%%%%%
%%%%%%%%%%%%%%%%%%%%%%%%%%%%%%%%%%%%
\section{Tree-level Form Factors}
\label{sec:treeFF}

In this section, we consider the general formalism for tree-level form factors. As we will see, the simple structure of Parke-Taylor like formula also exists in MHV form factors. Another simple class of tree building blocks are the minimal form factors, through which generic local operators can be represented by on-shell super spinor variables.  The tree-level results are building blocks of loop corrections in the unitarity construction, thus the simplicity at tree level implies the simplicity of form factors at loop level.

As a word about notation, through this review we will use $\phi, \psi, \bar\psi$ to represent on-shell asymptotic states and use $\Phi, \Psi, \bar\Psi$ to denote the (off-shell) elementary fields in the gauge invariant operators. They are related to each other by LSZ reduction.

%%%%%%%%%%%%%%%%%%%%%%%%%%%%%%
\subsection{Parke-Taylor-like form factors}

From the Feynman diagram point of view, there is no much difference between amplitudes and form factors. Besides the LSZ reduction for on-shell states, one also preforms Fourier transformation for the position $x$ of the local operator ${\cal O}(x)$. 
The locality of the operator is reflected in that the operator carries an off-shell momentum $q$.

Let us consider the form factor of operator ${1\over2}{\rm tr}(\Phi^2)$ and two external scalars. Following the definition \eqref{eq:def-FF}, one can easily compute it as:
\begin{align}
\hat F_{{1\over2}{\rm tr}(\phi^2),2} 
& = \int d^D x \, e^{-i q \cdot x} \prod_{j=1}^2 \int d^D x_j \, e^{i p_j \cdot x_j}  \partial_{x_j}^2 \langle \phi^{a_1}(x_1) \phi^{a_2}(x_2)|{1\over2}{\rm tr}(\Phi^2)(x) | 0\rangle \nonumber\\
& = (2\pi)^D\delta^{(D)}\Big(q-\sum_{i=1}^2 p_i\Big) {\rm tr}(T^{a_1} T^{a_2}) \times 1+ \big(\textrm{loop corrections} \big) \,,
\end{align}
in which the tree-level result contains a single trace color factor and the kinematic part is simply one. 

\begin{figure}[t]
\centerline{\includegraphics[height=3.5cm]{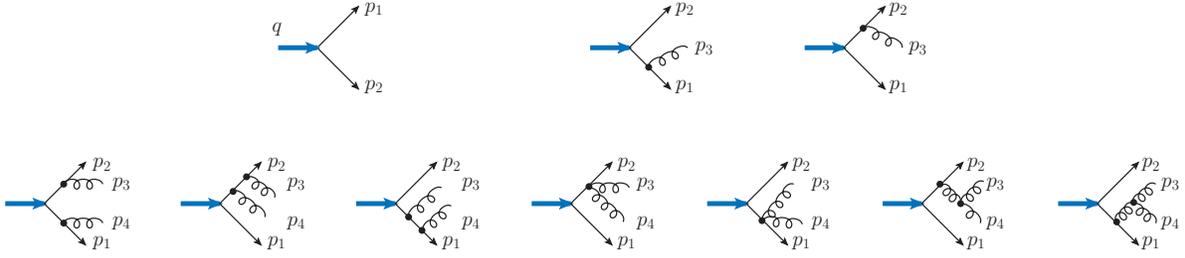} } 
\caption{Feynman tree diagrams for form factors of ${\rm tr}(\phi^2)$ up to four points.} 
\label{fig:Fss_tree}
\end{figure}

Like amplitudes, one can apply color decomposition for form factors. 
The planar form factors can be computed via planar Feynman diagrams with color-stripped Feynman rules. 
Note that since the operator is a color singlet, the color factor is independent of the position where the operator is inserted in the diagram.
In Figure~\ref{fig:Fss_tree}, we provides the planar Feynman diagrams for the form factors of ${1\over2}{\rm tr}(\Phi^2)$, up to four external legs.
For example, the three-point form factor contains two diagrams, which can be computed as:
\begin{equation}
F^{(0)}_{{1\over2}{\rm tr}(\phi^2),3}(1_\phi, 2_\phi, 3_g) = \delta^{(4)}(q-\sum_{i=1}^3 p_i) \sqrt{2} \left( {\varepsilon_3 \cdot p_1 \over s_{13}} - {\varepsilon_3 \cdot p_2 \over s_{23}} \right) \,.
\end{equation}

%%%%%%%%%%%%%%%%%%%%%%%%%%%%%%%%%%
%%%%%%%%%%%%%%%%%%%%%%%%%%%%%%%%%%

We can also consider \emph{helicity form factors} where the helicities of gluons are specified. The simplest class of form factors are the case of all plus-helicity gluons. Such planar form factors take the simple form \cite{Brandhuber:2010ad}:\footnote{The minus sign on the RHS is introduced such that the two-point form factor is $+1$. Similar minus signs also appear in the equations \eqref{eq:FF-superAll-MHV}--\eqref{eq:FNmaxMHV}.}
\begin{equation}
F^{(0)}_{{1\over2}{\rm tr}(\phi^2),n}(1_\phi, 2_\phi, 3^+, \ldots, n^+) = {-}\delta^{(4)}(q-\sum_i \lambda_i\tilde\lambda_i) {\langle 12 \rangle^2 \over \langle 12\rangle \langle 23\rangle \ldots \langle n 1 \rangle} \,.
\end{equation}
One can explicitly check this using the above Feynman diagram results. For general $n$-point, it can be proved using BCFW recursion relation. This formula is similar to the expression of the Parke-Taylor MHV amplitudes \eqref{eq:MHVamplitude}. The MHV form factors are more non-trivial in the sense that there is an off-shell momentum involved. In the context of Higgs amplitudes (which is equivalent to the form factor of ${\rm tr}(F^2)$ operator), the MHV structure was obtained in \cite{Dixon:2004za}.

In ${\cal N}=4$ SYM, besides a supersymmetric extension for the on-shell states, there also exists a supersymmetric extension for the local operators. One illuminating example is the chiral stress tensor supermultiplet, which contains the half-BPS operator ${1\over2}{\rm tr}(\Phi^{12}\Phi^{12})$ as a component (see e.g.~\cite{Eden:2011yp, Eden:2011ku}):
\begin{align}
{\cal T}(x, \theta^+) = e^{\theta_\alpha^+ Q_+^\alpha} {1\over2}{\rm tr}(\Phi^{12}\Phi^{12}) =  {1\over2}{\rm tr}(\Phi^{12}\Phi^{12}) + \ldots + (\theta^+)^4 {\cal L}\,.
\label{eq:calT-def}
\end{align}
In \eqref{eq:calT-def}, the index `$+$' = $1,2$ (and `$-$' = $3,4$),\footnote{We point out that the definition `$+$' = $1,2$ (and `$-$' = $3,4$) is different from the notation in \cite{Eden:2011yp, Eden:2011ku}. The choice here, for the purpose of simplicity, can be taken as an explicit projection of the Harmonic space  in \cite{Eden:2011yp, Eden:2011ku}.  For example, $\theta_\alpha^+ Q_+^\alpha= \theta_\alpha^1 Q_1^\alpha+\theta_\alpha^2 Q_2^\alpha$, and $(\theta^+)^4=\prod_{A=1}^2\prod_{\alpha=1}^2 \theta_\alpha^A $.}
and the $(\theta^+)^4$-component ${\cal L}$ is the chiral on-shell Lagrangian
\begin{equation}
{\cal L} = {\rm tr} \left[ {1\over2} F_{\alpha\beta}F^{\alpha\beta} - \sqrt{2} g_{\rm YM} \Psi^{\alpha A} [\Phi_{AB}, \Psi^B_\alpha] + {1\over8}g_{\rm YM}^2 [\Phi^{AB}, \Phi^{CD}] [\Phi_{AB}, \Phi_{CD}] \right] \,.
\label{eq:calL}
\end{equation}
The supersymmetric transformation with respect to the supercharge $Q$ is given in Appendix \ref{app:susy}.

Interestingly,  the super MHV form factors of the chiral stress tensor supermultiplet also take the Parke-Taylor like form \cite{Brandhuber:2011tv}
\begin{align}
{\cal F}_{{\cal T},n}^{(0),{\rm MHV}}(1,\ldots, n; q, \gamma_+) = & \int d^4 x \, d^4 \theta^+ \,e^{- i q\cdot x - \gamma_{+}^\alpha \theta^{+}_\alpha}  
\langle 1 \ldots n |{\cal T}(x, \theta^+) | 0\rangle \nonumber\\
= & - {\delta^{(4)}(q-\sum_i \lambda_i\tilde\lambda_i) \delta^{(4)}(\gamma_+ -\sum_i \lambda_i \eta_{+,i}) \delta^{(4)}(\sum_i \lambda_i \eta_{-i}) \over \langle 12\rangle \langle 23\rangle \ldots \langle n 1 \rangle} \,.
\label{eq:FF-superAll-MHV}
\end{align}
Like that $q$ is the momentum for the operator, the $\gamma_{+}$ plays the role of supermomentum for the super-operator. 
To obtain different components of super-operator, we can expand the fermionic delta function and take different coefficients in the $\gamma$ expansion. For example: 
\begin{align}
(\gamma)^4\textrm{ - term}: \qquad & {\cal F}_{{1\over2}{\rm tr}(\phi^2),n}^{(0),{\rm MHV}} = - {\delta^{(4)}(q-\sum_i \lambda_i\tilde\lambda_i) \delta^{(4)}(\sum_i \lambda_i \eta_{-i}) \over \langle 12\rangle \langle 23\rangle \ldots  \langle n 1 \rangle} \,, \\
(\gamma)^0\textrm{ - term}: \qquad & {\cal F}_{{\cal L},n}^{(0),{\rm MHV}} = - {\delta^{(4)}(q-\sum_i \lambda_i\tilde\lambda_i) \delta^{(8)}(\sum_i \lambda_i \eta_i^A) \over \langle 12\rangle \langle 23\rangle \ldots \langle n 1 \rangle} \,.
\label{eq:FF-super-component-MHV}
\end{align}
Using the supersymmetric form, one can show that the all negative helicity (i.e.~maximally non-MHV) form factors of ${\cal L}$ also take a very simple form \cite{Brandhuber:2011tv}:
\begin{equation}
{\cal F}_{{\cal L},n}^{(0),\textrm{N}^{\rm max}\textrm{MHV}} = - \delta^{(4)}(q-\sum_i \lambda_i\tilde\lambda_i){ (q^2)^2 \over [12] \ldots [ n 1 ]} (\eta_1)^4 \ldots (\eta_n)^4 \,. 
\label{eq:FNmaxMHV}
\end{equation}

The super form factors for a generalized class of half-BPS operators ${\cal T}_k$ containing ${\rm tr}(\phi^k)$ as a primary:
\begin{equation}
{\cal T}_k(x,\theta^+) = {1\over k} {\rm tr}[(\phi^{12})^k] + \ldots
\end{equation} 
have been studies in  \cite{Penante:2014sza},
and the MHV form factors also take compact form, for example for $k=3$, one has 
\begin{equation}
{\cal F}_{{\cal T}_3,n}^{(0),{\rm MHV}}(1,\ldots, n; q, \gamma_+) = {\cal F}_{{\cal T}_2,n}^{(0),{\rm MHV}}(1,\ldots, n; q, \gamma_+) \Big[ {1\over2}\sum_{i<j}^n \langle i j \rangle (\eta_i^1 \eta_j^2 - \eta_i^2 \eta_j^1) \Big] \,.
\end{equation}
MHV form factors of more generic operators were also studied using twistor formalism in \cite{Koster:2016loo, Chicherin:2016qsf}.

%%%%%%%%%%%%%%%%%%%%%%%%%%%%%%
\subsection{Minimal form factors of general operators}
\label{sec:mimiFF}
Unlike asymptotic on-shell states, local operators are off-shell and carry off-shell momenta. So at first sight, it is not obvious how on-shell formulation can be used to study them in a systematic way. This problem can be solved by using minimal form factor as a bridge, and spinor helicity formalism is useful to translate local operators to on-shell variables.
This connection between operators and the spinor helicity variables has been noticed in \cite{Beisert:2010jq, Zwiebel:2011bx}. The physical  interpretation of the this correspondence as form factors was realized in \cite{Wilhelm:2014qua} (see also \cite{Wilhelm:2016izi} for discussion).

%%%%%%%%%%%%%%%%%%
\subsubsection*{Local operators}
Let us first introduce general local gauge invariant operators. 
For simplicity, we also focus on the single-trace operators, which are given as the trace of field products:
\begin{align}
{\cal O}(x) = {\rm Tr}({\cal W}^{(m_1)}_1 {\cal W}^{(m_2)}_2 \ldots {\cal W}^{(m_L)}_L)(x) \,.
\end{align}
The covariant fields ${\cal W}_i$ can be any of the following field
\begin{align}
{\cal W}_i \in
\{ \Phi^{AB}\,, \ F^{\alpha\beta} \,, \ \bar F^{\dot\alpha\dot\beta} \,, \ \bar\Psi^{\dot\alpha A}\,, \ \Psi^{\alpha ABC} \} \,,
\end{align}
and each covariant field can be further dressed with powers of covariant derivatives
\begin{align}
{\cal W}^{(m)} := D^m {\cal W}\,, \qquad D_{\alpha\dot\alpha} {\cal W} = \partial_{\alpha\dot\alpha} {\cal W} - i g_{\rm YM} [A_{\alpha\dot\alpha}, \, {\cal W}] \,.
\end{align}

%%%%%%%%%%%%%%%%%%
\subsubsection*{Minimal form factors}

We define the \emph{length} $L$ of a given operator ${\cal O}$, as the number of fields ${\cal W}_i$ in the operator. 
The form factor of ${\cal O}$ is called {\it minimal}, if the number of the external legs match the length of the operator.   
Or equivalently, the minimal form factor is defined as ${\cal F}_{{\cal O}, L({\cal O})}$, such that ${\cal F}^{(0)}_{{\cal O}, L({\cal O})}\neq0$ while ${\cal F}^{(0)}_{{\cal O}, n}=0$ when $n<L({\cal O})$. 

In minimal form factors, there is a one-to-one correspondence between on-shell external particles and the fields ${\cal W}_i$ in the operator.
Let us explain this important point in more detail.
During the computation of the minimal tree form factor, every field in the operator is directly Wick contracted with an external field.
For example, let us consider the field strength $F_{\alpha\beta}$ in the operator ${\rm tr}(.. F_{\alpha\beta} ..)$. The $F_{\alpha\beta}$ needs to be contracted with a gluon field, say $A(x_j)$,
\begin{equation}
\contraction[1ex]
{\int d^4 x_j \, e^{i p_j \cdot x_j } \, \varepsilon^{(\pm)}_{\gamma\dot\gamma} \, \partial^2_{x_j} \langle 0| \ldots}
{A}
{{}^{\gamma\dot\gamma}(x_j) \ldots | {\rm tr}(\ldots }
{F}
\int d^4 x_j \, e^{i p_j \cdot x_j } \, \varepsilon^{(\pm)}_{\gamma\dot\gamma} \, \partial^2_{x_j} \langle 0| \ldots A^{\gamma\dot\gamma}(x_j) \ldots | {\rm tr}(\ldots F_{\alpha\beta} \ldots) |0\rangle \,,
\end{equation}
in which a LSZ reduction operation is also included for the gluon field $A(x_j)$.
After this computation, one can note that the field strength is effectively going through the following changes:
\begin{align}
F_{\alpha\beta} = {1\over2\sqrt{2}} \epsilon^{\dot\alpha\dot\beta} (\partial_{\alpha\dot\alpha} A_{\beta\dot\beta} - \partial_{\beta\dot\beta} A_{\alpha\dot\alpha}) 
& \xrightarrow[{\textrm{LSZ}}]{\textrm{Wick contr.}}
{\epsilon^{\dot\alpha\dot\beta} \over 2\sqrt{2}} \big( p_{j,\alpha\dot\alpha} \varepsilon^{(\pm)}_{j,\beta\dot\beta} - p_{j,\beta\dot\beta} \varepsilon^{(\pm)}_{j,\alpha\dot\alpha} \big) \\
& \xrightarrow{\ \varepsilon^{(-)} \ }
{\epsilon^{\dot\alpha\dot\beta} \over 2} \Big( \lambda_{j,\alpha} \tilde\lambda_{j,\dot\alpha} \frac{\lambda_{j,\beta} \tilde\xi_{\dot\beta}}{[\tilde\lambda_j\tilde\xi]} - \lambda_{j,\beta} \tilde\lambda_{j,\dot\beta} \frac{\lambda_{j,\alpha} \tilde\xi_{\dot\alpha}}{[\tilde\lambda_j\tilde\xi]} \Big) = \lambda_{j,\alpha} \lambda_{j,\beta} \,. \nonumber
\end{align}
In the last step, we have chosen the helicity of external gluon to be negative. 
With the choice of plus helicity gluon, a non-zero result can be obtained if one starts with $\bar F_{\dot\alpha\dot\beta}$. Similar procedure can be done for scalars and fermions. 

Therefore, through minimal form factor, the local operator is naturally translated in terms of spinor helicity variables:
\begin{equation}
\textrm{off-shell local field} \xrightarrow{\ \ \ \textrm{minimal form factor} \ \ } \textrm{on-shell spinor helicity quantity} \,.
\end{equation}
We summarize the correspondence in Table~\ref{tab:miniFF}.

%%%%%%%%%%%%%%%
\begin{table}[t]
\begin{center}
\caption{Correspondence between local operators and minimal form factors.
}
\label{tab:miniFF}
\vskip .5 cm
\begin{tabular}{ccc} 
\hline
\textrm{Fields in local operators} & & \text{On-shell legs in minimal form factors} \cr 
$\bar F^{{\dot \alpha}{\dot \beta}}$ & \quad  $\xrightarrow{\ \ \ g_+ \ \ }$ \quad  & $\tilde\lambda^{\dot\alpha} \tilde\lambda^{\dot\beta}$   \cr
$\bar\Psi^{\dot\alpha A}$ & \quad  $\xrightarrow{\ \ \ \bar\psi_{\dot\alpha A} \ \ }$ \quad & $\tilde\lambda^{\dot\alpha} \eta^A$ \cr
$\Phi^{AB}$ & \quad  $\xrightarrow{\ \ \ \phi_{AB} \ \ }$ \quad & $\eta^A\eta^B$  \cr
$\Psi^{\alpha ABC}$  & \quad  $\xrightarrow{\ \ \ \psi_{\alpha ABC} \ \ }$ \quad & $\lambda^\alpha \eta^A\eta^B\eta^C$ \cr
$F^{\alpha\beta}$ & \quad  $\xrightarrow{\ \ \ g_- \ \ }$ \quad & $\lambda^\alpha \lambda^\beta \eta^1\eta^2\eta^3\eta^4$ \cr
$D^{\alpha\dot\alpha}$ & \quad  $\xrightarrow{\ \ \  p_{\alpha\dot\alpha}  \ \ }$ \quad & $\lambda^\alpha \tilde\lambda^{\dot\alpha} $ \cr\hline
\end{tabular} 
\end{center}
\end{table}
%%%%%%%%%%%%%%%

Using the above rules, it is straightforward to obtain the minimal form factor for any given operator. For example:\footnote{We consider only the planar form factor and omit the single-trace color factor.}
\begin{align}
& {\rm tr}( F_{\alpha\beta} F^{\alpha\beta}) \rightarrow  \lambda_{1\alpha} \lambda_{1\beta} \lambda_2^{\alpha} \lambda_2^\beta  (\eta_1)^4(\eta_2)^4  + \textrm{cyc.perm.}(1,2)  = 2 \langle 1\,2\rangle^2(\eta_1)^4(\eta_2)^4\,, \\
& {\rm tr}(\bar F^{\dot\alpha}_{{~}\dot\beta} \bar F^{\dot\beta}_{{~}\dot\gamma} \bar F^{\dot\gamma}_{{~}\dot\alpha}) \rightarrow  \tilde\lambda_1^{\dot\alpha} \tilde\lambda_{1\dot\beta}  \tilde\lambda_2^{\dot\beta} \tilde\lambda_{2\dot\gamma} \tilde\lambda_3^{\dot\gamma} \tilde\lambda_{3\dot\alpha}  + \textrm{cyc.perm.}(1,2,3) = 3 [ 1\,2] [ 2\,3] [ 3\,1] \,,  \\
& {\cal O}_{\cal K} = \epsilon_{ABCD}{\rm tr}(\Phi^{AB}\Phi^{CD})  \rightarrow 2 \epsilon_{ABCD} \eta_1^A \eta_1^B \eta_2^C \eta_2^D \,.
\end{align}
Note that one needs to sum over all necessary cyclic permutations of external momenta to get the full form factor.

%%%%%%%%%%%%%%%%%%%
\subsubsection*{Sectors of operartors}

One can classify operators into different sectors. 
The simplest sector is the \emph{vacua} sector, which consists of only one type of scalars and no covariant derivatives. One example is ${\rm tr}((\Phi^{12})^2)$ which belong to the stress tensor supermultiplet.  These operators are protected BPS operators, in the sense that no UV renormalization is needed. 
The simplest non-BPS sector is the $SU(2)$ sector, where the operators consist of only two types of scalars, say $\{\Phi^{12}, \Phi^{13}\}$, and no covariant derivatives. 
Another simple sector is the $SL(2)$ sector, in which the operators contain only one type of scalars but allow arbitrary numbers of covariant derivatives.
One generalization of the $SU(2)$ sector is the $SO(6)$ sector, where all six scalar are allowed to appear in the operator. 
Another generalization is the $SU(3|2)$ sector, which contains three scalars and a single fermion: $\{ \Phi^{12}, \Phi^{23}, \Phi^{13}; \Psi^{123, \alpha}\}$.
We summarize the above sectors as
 \begin{align}
 {\rm Vacua}: & \ \   \{\Phi^{12}\} \,, \nonumber\\
 SU(2): & \ \   \{\Phi^{12}, \Phi^{13}\} \,, \nonumber\\
 SO(6): & \ \  \{\Phi^{AB}; \textrm{ for all } A,B\} \,,  \\
 SU(3|2) : & \ \ \{ \Phi^{12}, \Phi^{23}, \Phi^{13} , \Psi^{123, \alpha}\} \,, \nonumber\\
 SL(2): & \ \ \{\Phi^{12}, D^{\alpha \dot\alpha}\} \,. \nonumber
 \end{align}
Non-BPS operators have non-trivial anomalous dimensions through quantum corrections. 
The computation of operator anomalous dimensions is one central topic in the study of integrability of ${\cal N}=4$ SYM \cite{Beisert:2010jr}. 
We will apply form factors to compute anomalous dimensions in Section \ref{sec:FFandAD}.

%%%%%%%%%%%%%%%%%%%%%%%%%%%%%%%%%%%%
%%%%%%%%%%%%%%%%%%%%%%%%%%%%%%%%%%%%
\section{Sudakov Form Factors and Infrared Divergences}
\label{sec:SudakovFF}

In this section, we consider one of simplest classes of form factors, the Sudakov form factors, which have played important roles in the study of IR singularities of gauge theories \cite{Mueller:1979ih,Collins:1980ih,Sen:1981sd,Magnea:1990zb}. A Sudakov form factor is defined as the minimal form factor of a length-2 operators, or equivalently, a matrix element between a length-2 operator and two on-shell states. 

To be concrete, in this section we will focus on the Sudakov form factor of half-BPS operator ${\cal L}={1\over2}{\rm tr}(F_{\alpha\beta}^2)+\ldots$ in the stress tensor supermultiplet with two negative helicity gluons:
\begin{equation}
\label{eq:def-sudakovFF}
{\cal F}_{{\cal L},2}(1,2) = \int d^D x \, e^{-i q \cdot x} \langle 1_{g^-} 2_{g^-}|{\cal L}(x) | 0\rangle \,.
\end{equation}
We will first compute the one and two-loop Sudakov form factor using the unitarity method. Then we discuss color-kinematics duality and apply it (together with the unitarity method) to compute two and higher loop cases. 
Finally, we discuss the application of Sudakov form factor for understanding high loop IR structures.

%%%%%%%%%%%%%%%%%%%%%%%%
\subsection{Unitarity-cut method}
\label{sec:SudakovFFunitarity}

Let us start with the simple one-loop case, in which we will keep all details. 

\subsubsection*{The one-loop example}

%%%%%%%%%%%%%%%%%%%%%%%%%%%%%%%%%%
\begin{figure}[t]
\centerline{\includegraphics[height=2.cm]{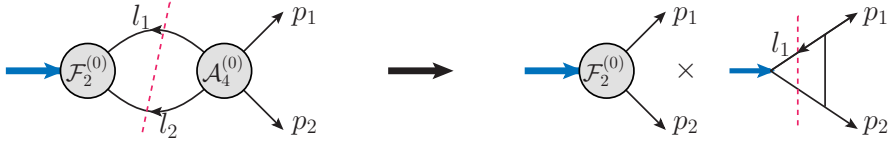} } 
\caption{Double-cut for the one-loop Sudakov form factor.} 
\label{fig:F2pt2cuts}
\end{figure}
%%%%%%%%%%%%%%%%%%%%%%%%%%%%%%%%%%

First of all, to determine one-loop Sudakov form factor, it is enough to consider the double cut, as shown in Figure~\ref{fig:F2pt2cuts}. The one-loop Sudakov form factor under this cut factorizes as a product of a two-point tree form factor and a four-point tree amplitudes:\footnote{As a simple illustration of the unitarity method, in this subsection we will consider planar form factors. For them we only need to consider planar cuts, where all building blocks are color-ordered quantities.}
\begin{equation}
{\cal F}_2^{(1)}(1,2) \big|_{s_{12}\textrm{-cut}} = \int d \textrm{PS}_2 \sum_\textrm{helicity of $l_i$} {\cal F}_2^{(0)}(-l_1,-l_2) {\cal A}_4^{(0)}(1, 2, l_2, l_1) \,,
\end{equation}
where
\begin{align}
& {\cal F}_2^{(0)}(-l_1,-l_2) = - \frac{\delta^{(8)}(\lambda_{l_1} \eta_{l_1} + \lambda_{l_2} \eta_{l_2})}{\langle l_1 l_2 \rangle \langle l_2 l_1 \rangle} \,, \\
& {\cal A}_4^{(0)}(1, 2, l_2, l_1) = i \frac{\delta^{(8)}(\lambda_1 \eta_1 + \lambda_2 \eta_2 + \lambda_{l_1} \eta_{l_1} + \lambda_{l_2} \eta_{l_2})}{ \langle 1 2 \rangle \langle 2 l_2 \rangle \langle l_2 l_1 \rangle \langle l_1 1 \rangle} \,,
\end{align}
and the cut integral measure becomes a two-particle phase space integral measure
\begin{align}
& d \textrm{PS}_2 = { d^D l_1 \over (2\pi)^D} 2\pi \delta_+\big( l_1^2 \big) 2\pi \delta_+\big( (l_1+p_1+p_2)^2 \big) \,.
\end{align}
One needs to sum over all possible internal states for the cut legs $l_1,l_2$. This can be achieved by the integration of Grassmann variables $\eta_{l_1}, \eta_{l_2}$ as \eqref{eq:etaIntegral}. One has 
\begin{align}
{\cal F}_2^{(1)}(1,2) \big|_{s_{12}\textrm{-cut}} = & \ \int d \textrm{PS}_2 \, \int d^4 \eta_{l_1} d^4 \eta_{l_2} {\cal F}_2^{(0)}(-l_1,-l_2) {\cal A}_4^{(0)}(1, 2, l_2, l_1) \nonumber\\
= &\ {\cal F}_2^{(0)}(1, 2) \,i \int d \textrm{PS}_2 \, \frac{\langle l_1 l_2 \rangle \langle 1 2\rangle}{\langle l_1 p_1\rangle \langle l_2 2\rangle} \nonumber\\
= &\ {\cal F}_2^{(0)}(1, 2) \,i \int d \textrm{PS}_2 \, \frac{-s_{12}}{(l_1+p_1)^2} \,,
\end{align}
where in the last step, the spinor products have been reorganized in terms of Lorentz products.
Now, by a reverse of the cut operation \eqref{eq:cutdef} with ${2\pi \delta_+(l^2) \rightarrow{i \over l^2}}$ in the measure factor $d \textrm{PS}_2$, one gets
\begin{align}
{\cal F}_2^{(1)}(1,2) \big|_{s_{12}\textrm{-cut}} = &\  {\cal F}_2^{(0)}(1, 2) \frac{(4\pi e^{- \gamma_{\text{E}}})^\epsilon}{(4\pi)^2} \, e^{\epsilon \gamma_{\text{E}}} \int {d^D l_1 \over i\pi^{D/2}} \, \frac{- s_{12}}{l_1^2 (l_1+p_1)^2 (l_1 + p_1 + p_2)^2} \Big|_{s_{12}\textrm{-cut}}  \nonumber\\ 
 = & \ {\cal F}_2^{(0)}(1, 2) \frac{(4\pi e^{- \gamma_{\text{E}}})^\epsilon}{(4\pi)^2}  (-s_{12})
\hskip -.2cm
\begin{tabular}{c}{\includegraphics[height=1.4cm]{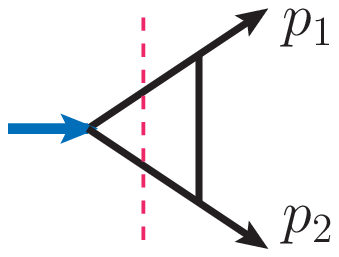} } \end{tabular}
\hskip -.4cm \,.
\end{align}
This shows that the cut integrand is equivalent to the cut of a triangle integral with $-s_{12}$ in the numerator.
(The scalar triangle integral $I_{\rm tri}$ is defined in \eqref{eq:scalartriangleintegral} in Appendix \ref{app:integrals}.)
The factor $\frac{(4\pi e^{- \gamma_{\text{E}}})^\epsilon}{(4\pi)^2}$ can be absorbed into the effective coupling constant $g^2$ defined as:
\begin{equation}
g^2 = \frac{g_{\rm YM}^2N_{\text{c}}}{(4\pi)^2}(4\pi e^{- \gamma_{\text{E}}})^\epsilon \,,
\label{eq:g2coupling}
\end{equation}
and we will omit this factor in later discussions.
Finally, we can remove the cuts and obtain
\begin{equation}
{\cal F}_2^{(1)}(1,2) = {\cal F}_2^{(0)}(1, 2) \times (-2 \, s_{12}) 
\begin{tabular}{c}{\includegraphics[height=1.5cm]{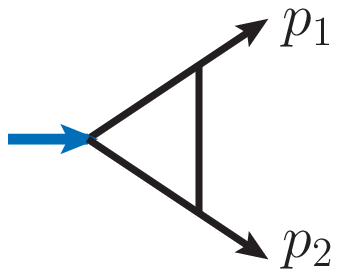} } \end{tabular} \hskip -.4cm .
\label{eq:Sudakov1loop}
\end{equation} 
A factor $2$ is added since one needs to sum over  the permutation of $p_1$ and $p_2$ to get the full planar form factor.

%
%%%%%%%%%%%%%%%%%%%%%%%%%%%%%%%%%%
\begin{figure}[t]
\centerline{\includegraphics[height=1.7cm]{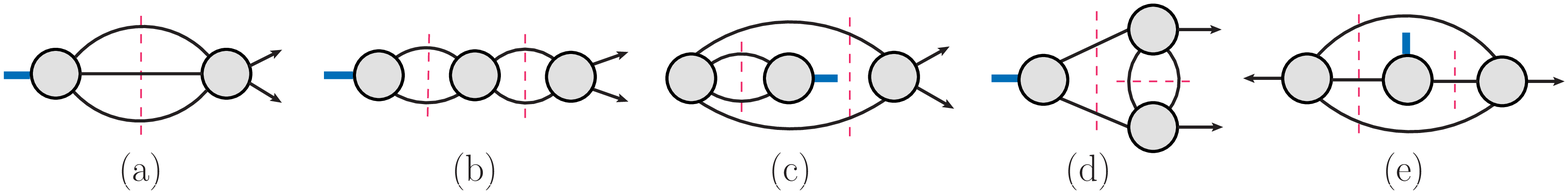}} 
\caption{A spanning set of cuts for the two-loop Sudakov form factor.} 
\label{fig:F2_2loop_Allcuts}
\end{figure}
%%%%%%%%%%%%%%%%%%%%%%%%%%%%%%%%%%

%%%%%%%%%%%%%%%%%%%%%%%%%%%%%%%%%%
\begin{figure}[t]
\centerline{\includegraphics[height=1.4cm]{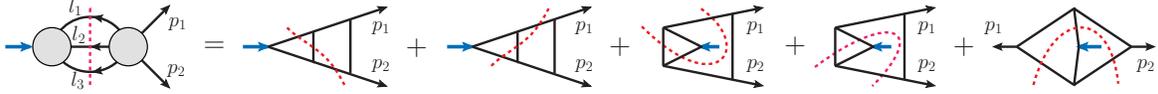} } 
\caption{The triple-cut for the two-loop Sudakov form factor.} 
\label{fig:triplecut_frombasis}
\end{figure}
%%%%%%%%%%%%%%%%%%%%%%%%%%%%%%%%%%

\subsubsection*{The two-loop example}

Next we consider the two-loop case. As in the one-loop case, the strategy is to consider the cuts and then use tree products to reconstruct the loop integrand. At two loops, since the topology is more complicated than the one-loop case, it is necessary to consider several different cuts. 
A spanning set of cuts that is enough to determine the two-loop result are shown in Figure~\ref{fig:F2_2loop_Allcuts}. 

Here we would like to emphasize a special feature of planar form factors that  does not occur in planar amplitudes. Because the operator is a color singlet, the operator leg $q$ can appear in the interior of the graph (such as cut (c) and (e) in Figure~\ref{fig:F2_2loop_Allcuts}), without changing the color planarity.  This means that the integrals of non-planar topologies can contribute in planar-color form factors. 

As an example, let us consider the triple cut (a) in Figure~\ref{fig:F2_2loop_Allcuts}. One can check that the product of tree quantities ${\cal F}_3^{(0)} \times {\cal A}_5^{(0)}$ are equivalent to the cut contribution from the basis integrals, as shown in Figure~\ref{fig:triplecut_frombasis}.   We can see that the last integral on the right hand side of Figure~\ref{fig:triplecut_frombasis} is of non-planar topology. Explicitly, Figure~\ref{fig:triplecut_frombasis} corresponds to the following equation
\begin{align}  
& \int \prod_{i=1}^3 d^4 \eta_{l_i}  \Big[  {\cal F}^{(0),{\rm MHV}}_3(-l_1, - l_2, - l_3) \, {\cal A}^{(0),{\rm NMHV}}_5(p_1, p_2, l_3, l_2, l_1)  \\ 
& \hskip 2.5cm +  {\cal F}^{(0),{\rm NMHV}}_3(-l_1, - l_2, - l_3) \, {\cal A}^{(0),{\rm MHV}}_5(p_1, p_2, l_3, l_2, l_1)  \Big]  \nonumber \\
= & {\cal F}_2^{(0)}(1,2) s^2_{12}\bigg( {1\over s_{l_1 l_2} s_{l_2 l_3} s_{2 l_3}} + {1\over s_{l_1 l_2} s_{l_2 l_3} s_{1 l_1}} + {1\over s_{l_2 l_3} s_{l_1 l_3} s_{1 l_1}} + {1\over s_{l_1 l_2} s_{l_1 l_3} s_{2 l_3}} + {1\over s_{l_1 l_3} s_{1 l_1} s_{2 l_3}} \bigg)  \,,  \nonumber
\end{align} 
where the expressions of tree form factor ${\cal F}^{(0)}_3$ can be found in \eqref{eq:FF-super-component-MHV}, \eqref{eq:FNmaxMHV}, and MHV amplitude ${\cal A}^{(0)}_5$ is given in \eqref{eq:AsuperMHV}. Using the spinor product relations, it is possible to transform the tree product expression as the Lorentz product expression on the right hand side which has a clear interpretation of integral basis. Alternative, as a check, one can simply expand both side in an independent spinor product basis as described in Appendix \ref{app:spinorbasis}.

The two-loop planar Sudakov form factor which is consistent with all cuts takes the following simple form:
\begin{equation}
{\cal F}_2^{(2)}(1,2) = {\cal F}_2^{(0)}(1, 2) \times s^2_{12} \left(
\ 2 \hskip -.2cm \begin{tabular}{c}{\includegraphics[height=1.2cm]{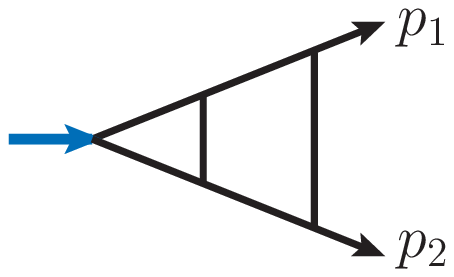} } \end{tabular} 
\hskip -.4cm +
\ 2 \hskip -.1cm \begin{tabular}{c}{\includegraphics[height=1.2cm]{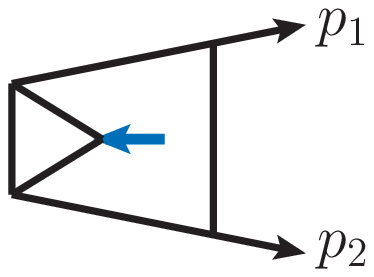} } \end{tabular}
\hskip -.4cm +
\hskip -.2cm \begin{tabular}{c}{\includegraphics[height=1.1cm]{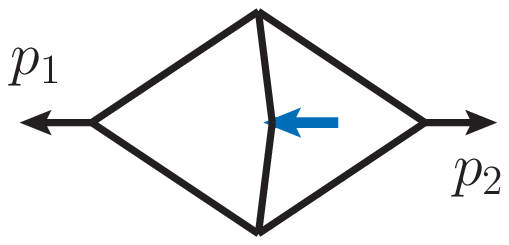} } \end{tabular}
\hskip -.4cm 
\right) ,
\label{eq:F2pt_2loop_viaunitarity}
\end{equation} 
where the factors $2$ in front the ladder integrals are from the sum of the permutation of external legs $p_1$ and $p_2$. 
This result was first obtained using Feynman diagram method \cite{vanNeerven:1985ja}. 
One may note that the first two integrals are mathematically identical. We separate them explicitly to indicate that in the planar cut we need to treat them differently.

%%%%%%%%%%%%%%%%%%%
\subsection{Color-kinematics duality and two-loop example}
\label{sec:CKSudakovFF}

Unitarity method at high loops can be complicated, due to the increasing number of topologies. 
It becomes much more efficient if one can combine unitarity method with other symmetry properties. The symmetry property can be used to first construct certain ansatz of the integrand, and then unitarity cuts serve as physical constraints and checks. If the ansatz can be solved to pass all possible unitarity checks, the obtained integrand is guaranteed to be the correct physical result.
One example where such strategy can be applied very efficiently is the planar amplitudes in ${\cal N}=4$ SYM, which enjoy the dual conformal symmetry \cite{Drummond:2006rz}. This property allows one to construct an ansatz of integrand as a linear combination of dual conformal basis integrals. Then the unitarity method can be used to fix the coefficients of the basis integrals. 

For form factors, however, because of the appearance of the color singlet operator, non-planar integrals are necessarily involved even in the (color) planar limit. This apparently breaks the dual conformal invariance, and computing non-planar topologies via unitarity is a non-trivial task. 
Luckily, the color-kinematics duality \cite{Bern:2008qj, Bern:2010ue} can be used to solve this problem for form factors \cite{Boels:2012ew,Yang:2016ear}.
The color-kinematics duality, by construction, involves full color factors, so it provides the great advantage of computing not only the planar but also non-planar parts.

Let us first describe the general strategy of using color-kinematics duality.  
As in the tree amplitude case in Section \ref{sec:CKduality}, the starting point is to represent an $L$-loop form factor as a sum over trivalent graphs $\Gamma_i$:
\begin{align}  
\hat{\cal F}_2^{(L)}  = {\cal F}_2^{(0)} \sum_{\sigma_2}  \sum_{\Gamma_i} \int \prod_j^L d^D l_j {1\over S_i} {C_i \, N_i \over \prod_a D_{i,a}} \,,
\label{eq:F2-cubicexpansion}
\end{align}
in which, except the numerators $N_i$, all other factors are fixed by the topologies:

\begin{itemize}

\item
The propagators $D_{i,a}$ are determined by the topologies. 

\item
The color factors $C_i$ are also determined by the topologies. They are defined as the product of $\tilde f^{abc}$ associated to each trivalent vertex. In form factors there is also a special vertex, the one connected to the operator-leg $q$, which is dressed with the factor $\delta_{ab} = {\rm Tr}(T^a T^b)$.  

\item
The $S_i$ are the symmetry factors, which account for the over-counting from the symmetries of the graph. In practice, $S_i$ can be computed (using e.g. Mathematica) by randomly shuffling the list of edges and then counting the number of inequivalent isomorphic maps.

\item
The sum of $\sigma_2$ takes into account the permutation of external on-shell legs. 

\end{itemize}

%%%%%%%%%%%%%%%%%%%%%%%%%%%%%%%%%%
\begin{figure}[t]
\centerline{\includegraphics[height=2.cm]{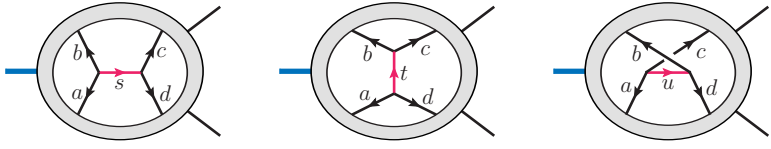} } 
\caption{Three loop topologies that are related by the colour-kinematics duality.} 
\label{fig:loopBCJ}
\end{figure}
%%%%%%%%%%%%%%%%%%%%%%%%%%%%%%%%%%

The color-kinematics duality plays the role of constraining the kinematic numerators $N_i$. 
For a trivalent graph, one can choose any propagator, which is not directly connected to the off-shell leg $q$. Consider the first figure in Figure~\ref{fig:loopBCJ}, and we pick up the propagator indicated by red color. We call this graph a $s$-channel graph, as indicated by the four-point tree subgraph. By changing the four-point tree topology, one can generate the other two graphs, as shown in Figure~\ref{fig:loopBCJ}, which are called $t$ and $u$-channel graphs.
Their color factors are
\begin{align} 
C_s =\tilde f^{a b s}\tilde f^{c d s} \prod \tilde{f} \,, \quad 
C_t =\tilde f^{b c t}\tilde f^{d a t} \prod \tilde{f} \,, \quad 
C_u =\tilde f^{a c u}\tilde f^{b d u}  \prod \tilde{f} \,,
\end{align}
where $\prod \tilde{f}$ are identical in all three factors.
Clearly, they satisfy the same Jacobi relation as the four-point tree amplitude:
\begin{align} 
C_s = C_t + C_u \,.
\end{align}

The assumption of color kinematics duality is to require that the kinematic numerators of the three graphs satisfy the same relation as color factors:
\begin{align} 
N_s= N_t + N_u\,.
\label{eq:CK-numerator-relation}
\end{align}
It is necessary to stress that, unlike the four-point tree amplitudes in Section~\ref{sec:CKduality}, the four legs $a,b,c,d$ are in general off-shell in Figure~\ref{fig:loopBCJ}. 
Thus a priori, such relations for the numerators are not guaranteed to be satisfied, and it is non-trivial that such a solution exists. 

By considering the relations from all possible propagators in all graphs, one can obtain a large set of equations for the kinematic numerators. After solving these equations, a small number of graphs can be identified, such that using their numerators, all other numerators can be obtained through the dual Jacobi relations. This small set of graphs will be called ``master graphs".
Then one only needs to construct an ansatz for the master graphs, thus reducing the complexity of the problem significantly.
This strategy has also been used successfully in constructing higher loop amplitudes in SYM, see e.g.~\cite{Bern:2010ue, Carrasco:2011mn, Bern:2012uf, Bern:2013qca, Bern:2014sna, Bern:2017yxu, Johansson:2017bfl, Bern:2017ucb, Kalin:2018thp}, as well as in pure YM \cite{Boels:2013bi, Bern:2013yya, Mogull:2015adi}. See \cite{Bern:2019prr, Carrasco:2015iwa} for more extensive review of the developments.

\subsubsection*{The two-loop example revisited}

Let us apply the above strategy to recompute the two-loop Sudakov form factor. This can be divided into following four steps:
%%%%%%%%%%%%%%%%%%%%%%%%%%%%%%%%%%
\begin{figure}[t]
\centerline{\includegraphics[height=1.3cm]{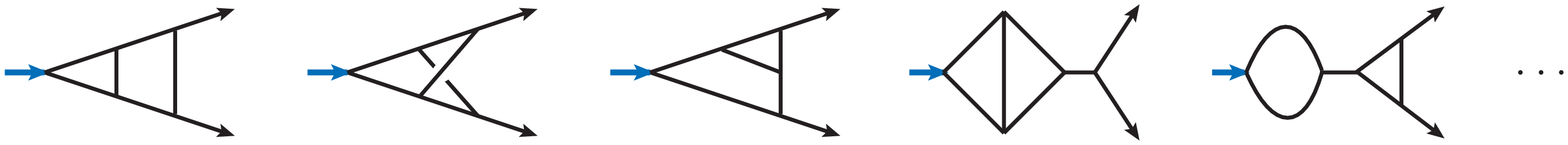} } 
\caption{Trivalent graphs of the two-loop Sudakov form factor.} 
\label{fig:F2_2loop_cubicgraph}
\end{figure}
%%%%%%%%%%%%%%%%%%%%%%%%%%%%%%%%%%

%%%%%%%%%%%%%%%%%%%%%%%%%%%%%%%%%%
\begin{figure}[t]
\centerline{\includegraphics[height=2.4cm]{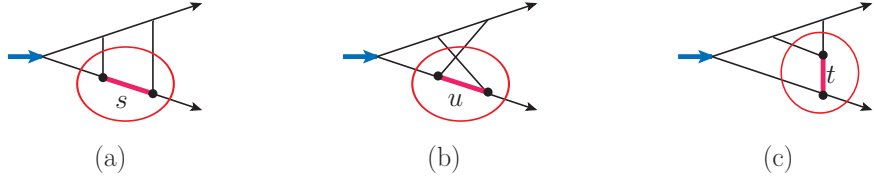} } 
\caption{Three graphs that are related by color-kinematics duality.} 
\label{fig:F2_2loop_BCJexample}
\end{figure}
%%%%%%%%%%%%%%%%%%%%%%%%%%%%%%%%%%

(1) {Generate cubic graphs.} The starting point is to generate all possible trivalent graphs, such as shown in Figure~\ref{fig:F2_2loop_cubicgraph}.
For ${\cal N}=4$ SYM, one may first exclude those graphs that are one-particle-reducible (1PR) or contain bubble/triangle subgraphs (no-triangle condition \cite{Bern:1994zx}). With these constraints, the allowed topologies are the first two graphs. In other words, we set the numerators of other topologies to be zero.

%%%%%%%%%%%%%%%%%

(2) {Find master graphs.} We then apply color-kinematics duality to generate all kinematic Jacobi relations,  which relate the numerators of different graphs. Consider the graph (a) in Figure~\ref{fig:F2_2loop_BCJexample} and the propagator in red color. 
As in Figure~\ref{fig:loopBCJ}, one can take it as the $s$-channel propagator. The corresponding $t$- and $u$-channel graphs are the graphs (b) and (c) in Figure~\ref{fig:F2_2loop_BCJexample}. Color-kinematics duality requires that the numerators satisfy:
\begin{equation}
N_{\rm a} = N_{\rm b} + N_{\rm c}  = N_{\rm b} \,,
\end{equation}
where we have used $N_{\rm c}=0$ since the graph contains a sub-triangle. 
We can choose graph (a) as the master graph.
Once we know the numerator of graph (a), we know the numerator of all graphs and thus the full integrand of the form factor.

(3) {Construct ansatz of master numerators.} We make an ansatz for the numerator of the master graph (a). 
Since the tree form factor is factored out in \eqref{eq:F2-cubicexpansion}, the numerator $N_a$ is expected to be a polynomial of Lorentz product of loop and external momenta: $\{l_i^2, l_i \cdot p_j, s_{12}\}$. 
For ${\cal N}=4$ SYM, one can further impose the constraint from the excellent UV property of the theory: for any $n$-point one-loop subgraph, the numerator should contain no more than $n-4$ powers of the loop momentum for that loop \cite{Bern:2012uf}, and if the one-loop subgraph is a form factor, the maximal power is $n-3$. Under this assumption, the only possible numerator of graph (a) is:
\begin{equation}
N_{\rm a} = s_{12}^2 \,,
\label{eq:Na2loop}
\end{equation}
up to a possible normalization factor which can be fixed by matching a simple unitarity cut.

(4) {Check unitarity cuts.} Finally, one needs to check against unitarity cuts in Figure~\ref{fig:F2_2loop_Allcuts}. As in the last subsection, it is straightforward to check that the numerator ansatz \eqref{eq:Na2loop} passes all checks.\footnote{Note that the ansatz depends on the full color factors. To check the ansatz, one can in principle consider the unitarity cuts with full color dependence. Alternative, it is more convenient to apply planar cuts as in Section \ref{sec:SudakovFFunitarity}. See \cite{Boels:2012ew} for the discussion on extracting the planar cut integrand from the ansatz. \label{footnote8}}

%%%%%%%%%%%%%%%
\begin{table}[t]
\begin{center}
\caption{The various factors for the two-loop Sudakov form factor.
\label{tab:F2_2loop}
}
\vskip .5 cm
\begin{tabular}{r | c | c | c} 
\hline
Graph &  Numerator factor  &  Color factor & Symmetry factor \cr \hline
(a) $\begin{matrix} ~ \\ ~   \end{matrix}$   &  $s_{12}^2$ & $  4 \, N_c^2 \, \delta^{a_1 a_2}  $ & 2 \cr \hline
(b) $\begin{matrix} ~ \\ ~   \end{matrix}$   & $s_{12}^2$ & $ 2 \, N_c^2 \, \delta^{a_1 a_2} $ & $4$ \cr \hline
\end{tabular} 
\end{center}
\end{table}
%%%%%%%%%%%%%%%

The results for various factors are summarized in Table~\ref{tab:F2_2loop}. 
The full form factor can be obtained as
\begin{equation}
 \hat{\cal F}_2^{(2)} =  {\cal F}_2^{(0)}  \,  \sum_{\sigma_2} \sum_{i={\rm a}}^{\rm b} {1\over S_i} \, C_i \, I_i   = N_c^2\, \delta^{a_1 a_2}\, {\cal F}_2^{(0)} \, \left( 4 \, I_a + I_b \right) \,,
 \end{equation}
 where the integrals $I_{a,b}$ are the planar and non-planar ladder graphs respectively.
This is indeed equivalent to the result we obtained in \eqref{eq:F2pt_2loop_viaunitarity}. Note that the computation with color-kinematics duality is for the form factor with full color dependence, and one can see that there is no non-planar correction in the two-loop case.

We would like to mention that in the above construction, we have made a few assumptions such as the no-triangle property and the good UV behavior of the numerators. These may not always work. In practice, one can always make these assumptions as a first try, and as long as the ansatz passes the unitarity checks, the result is correct. If not, one can then consider to relax these assumptions.

\subsection{Three-loop construction and beyond}

%%%%%%%%%%%%%%
\begin{figure}[t]
\centerline{\includegraphics[height=4cm]{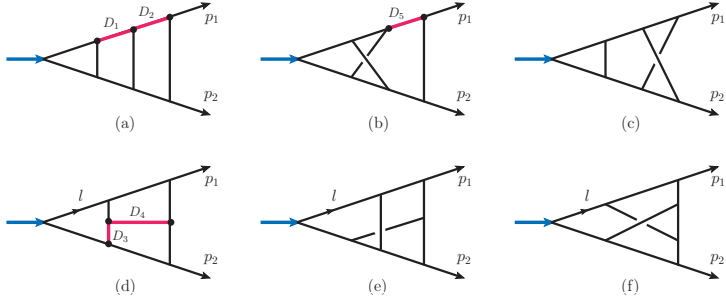} }
\caption{The six topologies for the three-loop Sudakov form factor.}
\label{fig:F2_3loop}
\end{figure}
%%%%%%%%%%%%%%

%%%%%%%%%%%%%%
\begin{figure}[t]
\centerline{\includegraphics[height=2cm]{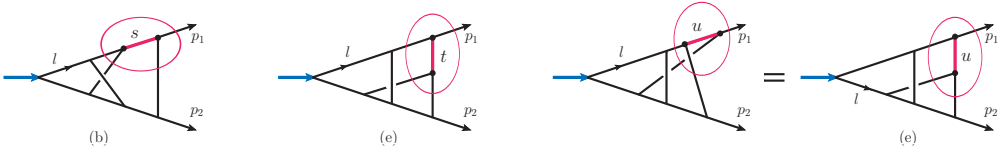}}
\caption{From Graph (b), one can perform $t$ and $u$ channel transformation for the propagator $s$, which generates two new graphs that has the same topology as graph (e). By color-kinematics duality, the numerator of graph (b) is equal to the sum of two numerators of graph (e), as given in \eqref{eq:3loopBCJ}.}
\label{fig:F2_3loop_BCJexample}
\end{figure}
%%%%%%%%%%%%%%

Next we consider the three-loop Sudakov form factor. 
We follow the same strategy as in the two-loop case:

(1) {Generate cubic graphs.} By excluding graphs that contain sub-bubble or sub-triangles, one can find there are only six topologies to consider, as shown in Figure~\ref{fig:F2_3loop}.

(2) {Find master graphs.} 
It turns out enough to consider the dual Jacobi relations associated to the propagators indicated by red color in  Figure~\ref{fig:F2_3loop}, and one obtains the following five relations
\begin{align}  
& N_{\rm a} \stackrel{\scriptscriptstyle D_1}{=} N_{\rm b}\,, \quad N_{\rm a} \stackrel{\scriptscriptstyle D_2}{=} N_{\rm c} \, ,  \quad N_{\rm d} \stackrel{\scriptscriptstyle D_3}{=} - N_{\rm e} \,, \quad N_{\rm d} \stackrel{\scriptscriptstyle D_4}{=} N_{\rm f} \, , 
\label{eq:3loopBCJ-simp} \\ 
& N_{\rm b}(p_1,p_2,l) \, \stackrel{\scriptscriptstyle D_5}{=} \, N_{\rm e}(p_1,p_2,l) + N_{\rm e}(p_1,p_2,p_1+p_2-l) \,.
\label{eq:3loopBCJ}
\end{align} 
The most non-trivial relation is the last relation \eqref{eq:3loopBCJ}. It can be understood using Figure~\ref{fig:F2_3loop_BCJexample}. Note that both $t$ and $u$-channel graphs are topologically equivalent to graph (e), therefore, their numerators are both given in terms of $N_{\rm e}$. 
By solving the equations \eqref{eq:3loopBCJ-simp}-\eqref{eq:3loopBCJ}, all numerators can be related to one single numerator, either $N_{\rm d}$ or  $N_{\rm e}$ or $N_{\rm f}$. We choose graph (d) as the master graph, since it is both planar and most symmetric. 
This will make it simpler to construct an ansatz for its numerator.

(3) {Construct ansatz of master numerators.} 
We construct an ansatz for $N_{\rm d}$ by applying the following constraints:
\begin{description}

\item[a)] 
With the power counting property discussed above \eqref{eq:Na2loop}, the numerator should depend linearly on the loop momentum $l$, since the topology contains a sub-box form factor.  A general ansatz can be given as
\begin{equation} 
N_{\rm d}^{\rm ansatz}(p_1,p_2,l) = (x_1 \, l \cdot p_1 + x_2 \, l \cdot p_2 + x_3 \,  p_1\cdot p_2) s_{12}^2  \, , 
\label{eq:ansatz-3loop-graphd}
\end{equation} 
which contains three parameters $x_i, i=1,2,3$.

\item[b)] 
We require further that the numerator satisfies the symmetry of the graph, which implies that it should be invariant under
\begin{equation} 
\{ p_1, p_2, l \} \quad \Longleftrightarrow \quad \{ p_2, p_1, p_1+p_2 - l \} \, , 
\end{equation} 
or more explicitly
\begin{equation} 
N_{\rm d} (p_1, p_2, l) = N_d(p_2, p_1, p_1+p_2 - l) \, . 
\label{eq:Ndsym}
\end{equation} 
Plugging the ansatz (\ref{eq:ansatz-3loop-graphd}) in \eqref{eq:Ndsym}, we obtain the relation
\begin{equation} 
x_2 = - x_1 \, .  
\end{equation} 

\item[c)] 
We consider further the simple constraint of the maximal cuts, where all propagators are taken on-shell. In such case the numerator should match the ``rung rule" numerator $(l - p_1)^2 s_{12}^2$ \cite{Bern:1998ug}:
\begin{equation}  
\big[ N_{\rm d} (p_1, p_2, l) - (l - p_1)^2 s_{12}^2 \big] \Big|_{\rm maximal~ cut} = 0  \, .  
\end{equation} 
This fixes the remaining two parameters as
\begin{equation} 
x_1 = -1, \qquad x_3 = -1 \, .  
\end{equation} 

\end{description}
Thus, by applying the above simple constraints we arrive at a unique solution for the master numerator:
\begin{equation} 
N_{\rm d}^{\rm ansatz} = [(p_2-p_1)\cdot l -p_1\cdot p_2] s_{12}^2  \ . 
\end{equation} 
Given this numerator, one can write down an ansatz for the full form factor.

%%%%%%%%%%%%%%
\begin{figure}[t]
\begin{center}
\includegraphics[height=2.5cm]{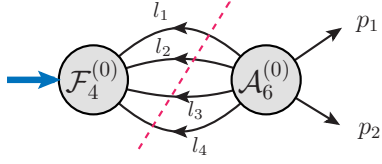}
\caption{A quadruple-cut for the three-loop Sudakov form factor.}
\label{fig:F2pt4cuts}
\end{center}
\end{figure}
%%%%%%%%%%%%%% . 
(4) {Check unitarity cuts.} We need to check if the above ansatz  satisfies the unitarity cuts:
\begin{equation}
\label{eq:unitarity-cut} 
{\cal F}^{(3)}|_{\rm cut} = \sum\textrm{cubic graphs} \Big|_{\rm cuts}  = \sum_{\rm helicities} \, F^{\rm tree} \prod_I A_I^{\rm tree} \,.
\end{equation}
One of the most constraining cuts is the quadruple-cut shown in Figure~\ref{fig:F2pt4cuts}. 
On one side, it is given by the product of tree results:
\begin{align}  
\int \prod_{i=1}^4 d^4 \eta_{l_i}  \Big[ & {\cal F}^{(0),{\rm MHV}}_4(-l_1, - l_2, - l_3, - l_4) \, {\cal A}^{(0),{\rm NNMHV}}_6(p_1, p_2, l_4, l_3, l_2, l_1) \nonumber \\ 
& +  {\cal F}^{(0),{\rm NMHV}}_4(-l_1, - l_2, - l_3, - l_4) \, {\cal A}^{(0),{\rm N}{\rm MHV}}_6(p_1, p_2, l_4, l_3, l_2, l_1) + \nonumber\\ 
& +  {\cal F}^{(0),\textrm{NNMHV}}_4(-l_1, - l_2, - l_3, - l_4) \, {\cal A}^{(0),{\rm MHV}}_6(p_1, p_2, l_4, l_3, l_2, l_1) \Big]  \, ,
\end{align} 
where the tree building blocks of non-MHV cases can be obtained using MHV rules \cite{Cachazo:2004kj}. On the other side, from the ansatz in terms of cubic integrals, one obtains a sum of $29$ cut diagrams. The equivalence of two sides provides a rather non-trivial check of the result.

%%%%%%%%%%%%%%%
\begin{table}[t]
\begin{center}
\caption{The factors for the three-loop Sudakov form factor.
\label{tab:F2_3loop}
}
\vskip .5 cm
\begin{tabular}{r | c | c | c} 
\hline
Graph &  Numerator factor  &  Color factor & Symmetry factor \cr \hline
(a) $\begin{matrix} ~ \\ ~   \end{matrix}$   &  $ s_{12}^3$ & $  8 \, N_c^3 \, \delta^{a_1 a_2}  $ & 2 \cr \hline
(b) $\begin{matrix} ~ \\ ~   \end{matrix}$   & $s_{12}^3$ & $ 4 \, N_c^3 \, \delta^{a_1 a_2} $ & $4$ \cr \hline
(c) $\begin{matrix} ~ \\ ~   \end{matrix}$   & $s_{12}^3$ & $4  \, N_c^3 \, \delta^{a_1 a_2}$ & $4$ \cr \hline
(d) $\begin{matrix} ~ \\ ~   \end{matrix}$   & $[(p_2-p_1)\cdot l -p_1\cdot p_2]s_{12}^2$ & $2  \, N_c^3 \, \delta^{a_1 a_2}$ & 2 \cr \hline
(e) $\begin{matrix} ~ \\ ~   \end{matrix}$   & $[- (p_2-p_1)\cdot l + p_1\cdot p_2]s_{12}^2$ & $2 \, N_c^3 \, \delta^{a_1 a_2}$ & 1 \cr \hline
(f) $\begin{matrix} ~ \\ ~   \end{matrix}$    & $[(p_2-p_1)\cdot l -p_1\cdot p_2]s_{12}^2$ & 0 & 2 \\ \hline
\end{tabular} 
\end{center}
\end{table}
%%%%%%%%%%%%%%%
The full form factor result can be finally obtained as
\begin{equation} 
\label{eq:F2-3loop-final}
\hat{\cal F}_2^{(3)} = {\cal F}_2^{(0)} \sum_{\sigma_2} \sum_{i={\rm a}}^{\rm e} {1\over S_i} \, C_i \, I_i \,,
\end{equation} 
where the various factors are summarized in Table~\ref{tab:F2_3loop}. 
Note that the graph (f) has zero color factor and therefore does not contribute to the final result of the form factor,
but it is necessarily involved in solving the Jacobi relations.
This result is consistent with the result in \cite{Gehrmann:2011xn} first computed using pure unitarity method.
With the aid of color-kinematics duality, the computation here is more straightforward. We can also note that there is no non-planar correction at three loops.

%%%%%%%%%%%%%%%%%%%%%%%%%
\subsubsection*{At higher loops}

%%%%%%%%%%%%%%%  TABLE   %%%%%%%%%%%%%%%%%%%%%
\begin{table}[t]
\centering
\caption{Number of the trivalent graphs and master graphs for Sudakov form factors up to 5 loops.
\label{tab:counting}
}
\vskip .5 cm
\begin{tabular}{l | c | c | c | c | c  } 
\hline
$L$ loops  		& \, $L$=1 \,  & \, $L$=2 \, & \, $L$=3 \, & \, $L$=4 \, & \, $L$=5 \,   \cr \hline 
\# of topologies   	&  1 & 2 & 6 & 34 & 306  \cr \hline 
\# of masters       	&  1 & 1 & 1 & 2 & 4  \cr \hline 
\end{tabular} 
\end{table}
%%%%%%%%%%%%%%%%%%%%%%%%%%%%%%%%%%%%%%%%%%

\begin{figure}[h]
\centerline{ \includegraphics[height=4cm]{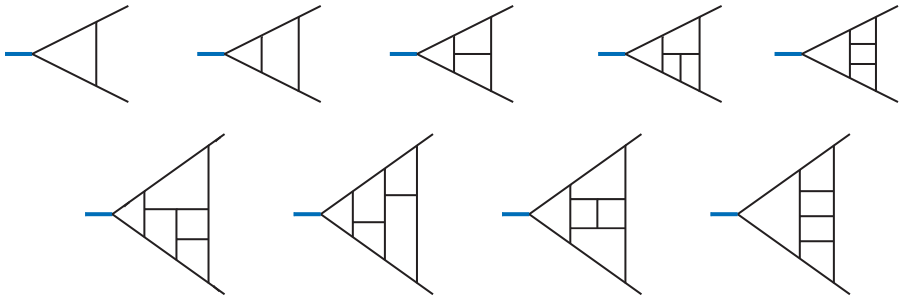} }
\caption{Master graphs for Sudakov form factors up to 5 loops.}
\label{fig:CKmasters}
\end{figure}

The above procedure has been successfully applied to construct Sudakov form factors at four and five loops \cite{Boels:2012ew,Yang:2016ear}.
We summarize the number of cubic graphs and master integrals in Table~\ref{tab:counting}. The corresponding master graphs are shown in Figure~\ref{fig:CKmasters}. We would like stress that the number of masters is 2 at four loops and 4 at five loops, which are remarkably small numbers comparing to the total numbers of cubic graphs that contribute to the results. 
In this way, the use of color-kinematics duality reduces the very non-trivial high loop construction to a much simpler problem.  Explicit duality satisfied solutions for full four- and five-loop Sudakov form factor are given in  \cite{Boels:2012ew,Yang:2016ear}.

%%%%%%%%%%%%%%
%%%%%%%%%%%%%% .

%%%%%%%%%%%%%%%%%%%%%%%%
\subsection{Infrared structure and non-planar corrections}
\label{sec:IRstructure}

Sudakov form factor plays a key role in the study of IR singularities of gauge theories \cite{Mueller:1979ih,Collins:1980ih,Sen:1981sd,Magnea:1990zb}. For example, it is an essential piece of information for the proposal of the BDS ansatz for amplitudes in ${\cal N}=4$ SYM \cite{Bern:2005iz}. 

Here we review the general structure of Sudakov form factor in ${\cal N}=4$ SYM. Since the operator we consider is the half-BPS operator, there is no UV divergences. (The non-BPS operators will be considered in next section.) We normalize the form factor as
\begin{equation}
{\cal F}_2^{(l)} = {\cal F}_2^{(0)} f = {\cal F}_2^{(0)} \sum_{l} g^{2 {l}} f^{(l)} \,,
\end{equation}
where the coupling constant $g^{2}$ is defined in \eqref{eq:g2coupling}.

The logarithm of the loop correction function $f$ takes the following structure \cite{Bern:2005iz}:
\begin{align}
\log f = \sum g^{2l} (\log f)^{(l)}  = - \sum_{{l}} g^{2l}(-q^2)^{-l \epsilon}  \bigg[ \frac{\gamma_{\textrm{cusp}}^{({l})} }{(2 {l} \epsilon)^2} + \frac{{\cal G}_{\textrm{coll}}^{({l})} }{2 {l} \epsilon} + {\rm Fin}^{(l)} \bigg] + {\mathcal O}\left(\epsilon\right) \,,
\label{eq:logfstructure}
\end{align}
where $\gamma_{\textrm{cusp}}$ is the cusp (soft) anomalous dimension \cite{Korchemsky:1988si}, and ${\cal G}_{\textrm{coll}}$ is the collinear anomalous dimension (see e.g.~\cite{Cachazo:2007ad, Dixon:2017nat}). 
The remarkable property of the IR divergences is that after taking the logarithm, the Sudakov form factor contains at most double pole in $\epsilon$.
We can perform a simple check of this property at one and two loops based on our previous computation. Using the integrand results \eqref{eq:Sudakov1loop} and \eqref{eq:F2pt_2loop_viaunitarity}, and together with the integral expressions in Appendix \ref{app:integrals}, one obtains
\begin{align}
(\log f)^{(1)} &= f^{(1)} =  (-2 s_{12}) I_3^{(1)} = (-s_{12}^2)^{-\epsilon} \Big[ - {2\over\epsilon^{2}} + {\cal O}(\epsilon^{0}) \Big] \,, \\
(\log f)^{(2)} &= f^{(2)} - {1\over2} (f^{(1)})^2 =  s_{12}^2 \Big(4 I_{\rm PL}^{(2)} + I_{\rm CL}^{(2)}\Big) - {1\over2} \Big( (-2 s_{12}) I_3^{(1)} \Big)^2 \nonumber\\
&  = (-s_{12}^2)^{-2\epsilon} \Big[  {\zeta_2 \over\epsilon^{2}} + {\zeta_3 \over\epsilon} + {\cal O}(\epsilon^{0}) \Big] \,.
\end{align}
All higher order poles in $\epsilon$ cancel in the two-loop case. Comparing with \eqref{eq:logfstructure}, one gets 
\begin{align}
\gamma_{\rm cusp}^{(1)} & = 8 \,, \qquad \qquad G_{\rm coll}^{(1)} = 0 \,, \label{eq:1loopcuspcoll} \\
\gamma_{\rm cusp}^{(2)} &= - 16 \zeta_2  \,, \qquad G_{\rm coll}^{(2)} = - 4 \zeta_3 \,. \label{eq:2loopcuspcoll} 
\end{align}

%%%%%%%%%%%%%%%%%%%%%%%%%%%%%%%%%%%%%%
\subsubsection*{Non-planar violation of the Casimir scaling conjecture}
Let us discuss an important fact about the non-planar color factors.
Up to three loops, the Sudakov form factor has no non-planar corrections: the color factors of $l$-loop Sudakov form factor are simply $N_c^l$ for $l=2,3$, as we can see from Table~\ref{tab:F2_2loop} and \ref{tab:F2_3loop}. New color structure appears at four loops. This is mainly because that the four-loop topology is the first case that contains eight cubic vertices, see the four-loop graphs in Figure~\ref{fig:CKmasters}. Thus its color factor involves a product of eight $f^{abc}$'s. 

This allows the appearance of a new quartic Casimir invariant:
\begin{equation}
d_{44} = d_A^{abcd}d_A^{abcd}/N_A \,,
\end{equation}
where
\begin{equation}
d_A^{abcd} = \frac{1}{4!} [ f^{\alpha a \beta} f^{\beta b \gamma} f^{\gamma c \delta} f^{\delta d \alpha} + {\text{perms.}(a,b,c,d)} ] \, . 
\end{equation}
For $SU(N_c)$ gauge group, $N_A = N_c^2-1$, and using \eqref{eq:fabc-def} and \eqref{eq:colorcontra}, one can obtain $d_{44} = N_c^2\, (N_c^2+36)/24$. (More details of computing color factors are given  in Appendix \ref{app:coloralgebra}.) Therefore, the fourth loop order is the leading order where the form factor can acquire a non-planar correction in the large $N_c$ expansion. 
See e.g.~\cite{Boels:2012ew} for a discussion for higher loop cases.

Although the planar cusp anomalous dimension in principle can be computed to all order via integrability \cite{BES06}, the non-planar corrections so far can be accessed only through explicit perturbative computation.
Given the integrand constructed by unitarity and color-kinematics duality \cite{Boels:2012ew}, it is still a very challenging task to compute the four-loop integrals. The four-loop integrand reduction for the ${\cal N}=4$ Sudakov form factor was achieved in \cite{Boels:2015yna}.  
The computation of integrals was realized by expanding the integrand in a set of uniform transcendentality basis \cite{Boels:2017skl, Boels:2017ftb}. We will not discuss the details of these computation since they go beyond the on-shell methods we focus in this review. We refer interested readers to the above original references.  See also \cite{Henn:2016men, vonManteuffel:2016xki, Lee:2016ixa, Lee:2017mip, vonManteuffel:2019wbj, vonManteuffel:2019gpr} for the ongoing efforts of computing four-loop Sudakov form factors beyond ${\cal N}=4$ SYM.

Let us quote the four-loop non-planar cusp and collinear anomalous dimensions \cite{Boels:2017ftb}:
\begin{equation}
\label{eq:results}
\gamma_{\textrm{cusp, NP}}^{(4)}  = -3072\times( 1.60 \pm 0.19 ) \frac{1}{N_c^2} \,, \qquad {\cal G}_{\textrm{coll, NP}}^{(4)}  = -384\times( -17.98 \pm 3.25 ) \frac{1}{N_c^2} \, .
\end{equation}
The concrete non-zero results show explicitly that the proposed Casimir scaling behavior (see e.g.~\cite{Becher:2009cu, Gardi:2009qi, Becher:2009qa, Dixon:2009ur}) is violated by the non-planar quartic Casimir corrections starting from four loops.
The result of non-planar cusp anomalous dimension in \eqref{eq:results} has been confirmed by an independent computation with higher precision \cite{Henn:2019rmi}, and its analytic form has been obtained recently first by a computation based on Wilson loop \cite{Henn:2019swt} and then by an independent computation based on the Sudakov form factor \cite{Huber:2019fxe}:
\begin{equation}
\label{eq:cuspAD4loopAnalytic}
\gamma_{\textrm{cusp, NP}}^{(4)}  = \frac{2}{N_c^2} \bigg[-576 \zeta _3^2-\frac{11904}{35} \zeta _2^3\bigg] \, .
\end{equation}
See e.g.~\cite{Moch:2017uml, Grozin:2017css, Moch:2018wjh, Lee:2019zop, Bruser:2019auj} for further computation of quartic Casimir corrections beyond ${\cal N}=4$ SYM.

In the next section, we will consider form factors with general non-BPS operators, which will contain both IR and UV divergences. Their IR divergences can be determined by the universal structure of Sudakov form factor that we have discussed in this section.

%%%%%%%%%%%%%%%%%%%%%%%%%%%%%%%%%%%%
%%%%%%%%%%%%%%%%%%%%%%%%%%%%%%%%%%%%
\section{Generic Form Factors and Anomalous Dimensions}
\label{sec:FFandAD}

In this section, we consider loop form factors of general non-protected operators. 
Scattering amplitudes in ${\cal N}=4$ SYM are UV finite since the beta function is zero.
On the other hand, the form factors and  correlation functions of general non-protected operators contain UV divergences and require renormalization. 
This is a general feature of CFT, where the scaling dimensions of operators can differ from their canonical dimensions:
\begin{equation}
\Delta_{\cal O} = \Delta_{{\cal O},0} + \gamma_{\cal O}(g) \,,
\end{equation}
where $\Delta_{{\cal O},0}$ is the (classical) canonical dimension, and $\gamma_{\cal O}(g)$ is the (quantum) \emph{anomalous dimension}.
The computation of operator anomalous dimensions is one central topic in studying the  AdS/CFT correspondence \cite{Maldacena:1997re} and the integrability of ${\cal N}=4$ SYM \cite{Beisert:2010jr}. 

One of our focus here is to apply on-shell methods to compute UV divergences such as the anomalous dimensions.
Below we first explain the above picture in the context of form factors. Then we will discuss explicit examples in the SO(6) and SL(2) sectors.

%%%%%%%%%%%%%%%%%%%%%%%%%%%%%%%%
\subsection{Renormalization of form factors}
\label{sec:renormalizationFF}

The operator renormalization can be carried out as
\begin{align}
\mathcal{O}_I^\text{ren} & = \sum_J \mathcal{Z}_I^{~J} \, \mathcal{O}_J^\text{bare} \,,  
\end{align}
where ${\cal Z}$ is the renormalization constant. Since form factor is linear in the operator,  as defined in \eqref{eq:def-FF}, a renormalized form factor is related to the bare one as
\begin{align}
{\cal F}_{\mathcal{O}_I}^\text{ren} & = \sum_J \mathcal{Z}_I^{~J} \, {\cal F}_{\mathcal{O}_J}^\text{bare} \,.
\label{eq:Fren-def}
\end{align}

It is worth to stress that the renormalization constant ${\cal Z}$ is in general a matrix. This is related to an important fact in QFT, the so-called operator mixing.
In the correlation function picture, the operator mixing can be understood by computing two-point function of two different operators $\langle {\cal O}_I(x) {\cal O}_J(x) \rangle$. 
In the framework of form factors, one can compute form factors of a given operator ${\cal O}_I$ with all possible external state configurations. According to the discussion in Section \ref{sec:mimiFF}, a given external state configuration (labelled by $J$) can be mapped to a  tree minimal form factors ${\cal F}_{\mathcal{O}_J}^{(0)}$ of operators ${\cal O}_J$.\footnote{We point out that operator mixing can also happen between the operators of different lengths. In the case that ${\cal O}_J$ has shorter length than  ${\cal O}_I$, the external states will be mapped to a non-minimal form factor of  ${\cal O}_J$. \label{footnote9}}  Therefore, different choices of external states will encode the operator mixing between ${\cal O}_I$ and ${\cal O}_J$.

Let us express the $l$-loop bare form factor as
\begin{align}
{\cal F}_{\mathcal{O}_I}^{(l),\text{bare}} & = \sum_J (\mathcal{I}^{(l),\text{bare}})_{I}^{~J} \, {\cal F}_{\mathcal{O}_J}^{(0)} \,,
\label{eq:Fbare-def}
\end{align}
where a sum over different external state configurations $J$ is understood.
Note that the loop correction functions $\mathcal{I}^{(l),\text{bare}}$ should be also considered as a matrix.
Similarly, the renormalized form factor is
\begin{align}
{\cal F}_{\mathcal{O}_I}^{(l),\text{ren}} & = \sum_J (\mathcal{I}^{(l),\text{ren}})_{I}^{~J} \, {\cal F}_{\mathcal{O}_J}^{(0)} \,.
\end{align}
Plugging \eqref{eq:Fbare-def} into \eqref{eq:Fren-def}, and expanding up to one and two-loop orders, one can obtain the relations:
\begin{align}
(\mathcal{I}^{(1),\text{ren}})_{I}^{~J} & =  (\mathcal{I}^{(1),\text{bare}})_{I}^{~J} + (\mathcal{Z}^{(1)})_{I}^{~J} \,, 
\label{eq:Fren1loop} \\ 
(\mathcal{I}^{(2),\text{ren}})_{I}^{~J} & =  (\mathcal{I}^{(2),\text{bare}})_{I}^{~J} + (\mathcal{Z}^{(2)})_{I}^{~J} + \sum_{K} (\mathcal{Z}^{(1)})_{I}^{~K} (\mathcal{I}^{(1),\text{bare}})_{K}^{~J}  \,.
\label{eq:Fren2loop}
\end{align}
Therefore, from bare form factors, one can obtain renormalization constants by requiring that UV divergences cancel systematically.

Given the renormalization constants ${\cal Z}$, the anomalous dimension matrix $\gamma$, also called the dilatation operator $\delta \mathds{D}$, can be obtained as
\begin{equation}
\gamma(g) =\delta \mathds{D}(g) = \mu \frac{d}{d \mu}\log {\cal Z} = 2\epsilon g^2 \frac{\partial}{\partial g^2} \log {\cal Z}=\sum_{l=1 }^\infty g^{2l} \mathds{D}^{(l)} .
\end{equation}
Expanding at one and two loops, one can obtain the explicitly relations:
\begin{equation}
\mathds{D}^{(1)} =  2\epsilon {\cal Z}^{(1)} \,, \qquad
\mathds{D}^{(2)} = 4\epsilon^2 \big[ {\cal Z}^{(2)} -{1\over2} ({\cal Z}^{(1)})^2 \big]  \,.
\label{eq:DvsZ}
\end{equation}
By diagonalizing the anomalous dimension matrix, one can also obtain the eigen-operators (which are eigenstates) and their corresponding eigenvalues.

The above discussion is the standard procedure of quantum field theory. 
The main point here is that one can use modern on-shell methods to compute form factors, and then also compute the UV information. Such a strategy has been used in  \cite{Wilhelm:2014qua, Nandan:2014oga, Loebbert:2015ova, Brandhuber:2016fni, Loebbert:2016xkw, Caron-Huot:2016cwu}.
We will review this strategy with explicit examples below. 
For simplicity, we will restrict ourselves to the planar limit.

%%%%%%%%%%%%%%%%%%%%%%%%%%%%%%%%
\subsection{SO(6) sector at one-loop}
\label{sec:SO6AD}

The first example we consider is the SO(6) sector, where the operators contain only scalar fields $\Phi_I, I = 1, \ldots, 6$ (or in the SU(4) notation, $\Phi_{AB}$):
\begin{equation}
{\rm tr}(\ldots \Phi_I \Phi_J \Phi_K \ldots) \,.
\end{equation} 
This sector played an important role in the discovery of the integrability of ${\cal N}=4$ SYM: the one-loop dilation operator in this sector was found to be identical to an integrable Heisenberg spin chain Hamiltonian \cite{Minahan:2002ve}:
\begin{equation}
(\mathds{D}^{(1)})_{\rm SO(6)} = \mathds{H}_{\rm SO(6)} = \sum_i 2(\mathds{1}- \mathds{P})_{i\,i+1}+ \mathds{T}_{i\,i+1} \,.
\label{eq:Hso6}
\end{equation}
In this section, we will reproduce this result using form factors with on-shell methods. 

We note that at one loop, the interaction only involves two fields in the operator and they are adjacent in the planar limit. Without loss of generality, we can focus on two adjacent fields $\Phi_I$, $\Phi_J$ in the operator and consider the contribution where these two fields are involved in the one-loop interaction. 
We will call this contribution as a density contribution, denoted as $F_{{\cal O}_{\Phi_I \Phi_J}}^{(1)}$. The full one-loop planar form factor is given by summing over densities from all adjacent two fields. 

\begin{figure}[t]
\centerline{ \includegraphics[height=1.8cm]{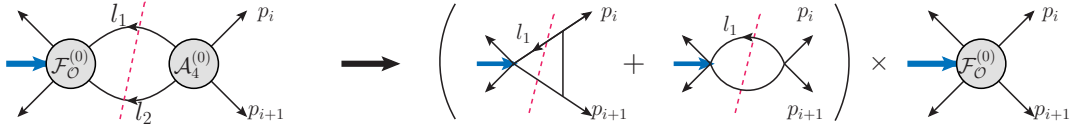} }
\caption{Double-cut for one-loop form factor.}
\label{fig:F2pt2cutsSO6SL2}
\end{figure}

As discussed in section \ref{sec:SudakovFFunitarity}, to determine the one-loop correction, it is enough to consider the double cut shown in Figure~\ref{fig:F2pt2cutsSO6SL2}.
%%%%%%%%%%%%%%
%%%%%%%%%%%%%% 
The cut form factor density is
\begin{equation}
F_{{\cal O}_{\Phi_I \Phi_J}}^{(1)} \big|_{\textrm{cut}} = \int d \textrm{PS}_2 \, F_{{\cal O}_{\Phi_I \Phi_J}}^{(0)}(l_1^{\phi_I}, l_2^{\phi_J}) \times A_4^{(0)}(p_i^{\phi_L}, p_{i+1}^{\phi_K}, l_2^{\phi_J}, l_1^{\phi_I})  \,.
\label{eq:FcutSO6}
\end{equation}
As explained above, we only focus on the contribution from two adjacent fields $\Phi_I$ and $\Phi_J$ in the operator, which is indicated by ${{\cal O}_{\Phi_I \Phi_J}}$. The cut legs $l_1$ and $l_2$ correspond to scalar particles $\phi_I$ and $\phi_J$ respectively.
The tree form factor is
\begin{equation}
F_{{\cal O}_{\Phi_I \Phi_J}}^{(0)}(l_1^{\phi_I}, l_2^{\phi_J}) = 1 \,,
\end{equation}
while the four-point amplitudes were computed in \eqref{eq:A4scalar1} as\footnote{One may notice that there are other non-zero tree amplitudes where $p_i, p_{i+1}$ are not scalars, such as $A_4^{(0)}(p_i^{g_+}, p_{i+1}^{g_-}, l_2^{\phi_J}, l_1^{\phi_I})$. They will contribute non-zero cut integrands. However, such non-scalar contributions are all zero after integration. Therefore, the SO(6) sector is closed by itself.  See \cite{Nandan:2014oga} for more discussion on this point.}
\begin{equation}
A_4^{(0)}(p_i^{\phi_L}, p_{i+1}^{\phi_K}, l_2^{\phi_J}, l_1^{\phi_I}) =  - \left( {s_{ii+1} \over s_{i l_1} } + 1 \right) \mathds{1} + \mathds{P} - \left(1 + {s_{i l_1} \over s_{ii+1}} \right) \mathds{T}  \,.
\end{equation}
In this computation, we choose to use the bosonic tree results without Grassmman $\eta$ variables. A computation with super tree quantities can be found in \cite{Nandan:2018hqz} (see also \cite{Loebbert:2015ova} for the $SU(2)$ sector).
Plugging these tree results in \eqref{eq:FcutSO6}, the cut integrand can be expressed as
\begin{equation}
F_{{\cal O}_{\Phi_I \Phi_J}}^{(1)} \big|_{\textrm{cut}} = 
\bigg\{
( - s_{i i+1} \mathds{1} ) 
\hskip -.2cm
\begin{tabular}{c}{\includegraphics[height=1.5cm]{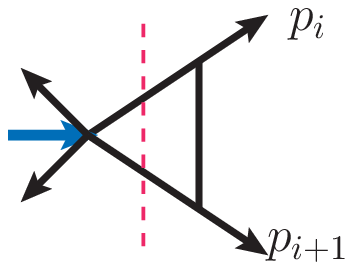} } \end{tabular}
\hskip -.7cm + \left[ - \mathds{1} + \mathds{P}  - \left(1 + {s_{i l_1} \over s_{ii+1}} \right) \mathds{T} \right] 
\hskip -.3cm
 \begin{tabular}{c}{\includegraphics[height=1.5cm]{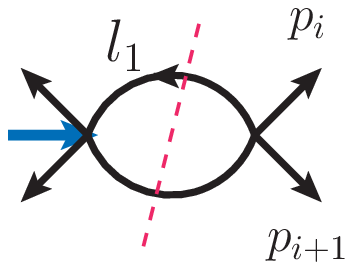} }\end{tabular}
 \hskip -.4cm \bigg\}
\cdot F_{{\cal O}_{\Phi_L \Phi_K}}^{(0)}  \,.
\end{equation}
After removing the cuts and using the relation of bubble integral $I_{\rm bub}[s_{il_1}] =  -{s_{i i+1} \over 2} I_{\rm bub}[1]$ (which is easy to obtain using PV reduction), one gets:
\begin{equation}
F_{{\cal O}_{\Phi_I \Phi_J}}^{(1)}  = (I_i^{(1)})_{\Phi_I \Phi_J}^{\Phi_L \Phi_K} \cdot F_{{\cal O}_{\Phi_L \Phi_K}}^{(0)} \,,
\end{equation}
where
\begin{equation}
(I^{(1)})_{\Phi_I \Phi_J}^{\Phi_L \Phi_K}  = ( - s_{i i+1} \mathds{1} ) 
\begin{tabular}{c}{\includegraphics[height=1.5cm]{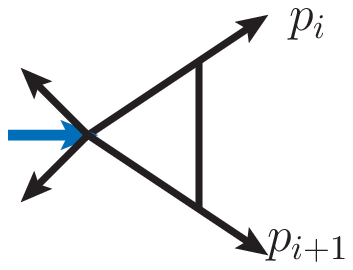} } \end{tabular}
\hskip -.5cm + \ \left( - \mathds{1} + \mathds{P}  - {1\over 2} \mathds{T} \right) 
 \begin{tabular}{c}{\includegraphics[height=1.5cm]{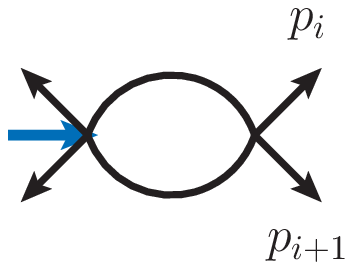} }\end{tabular} 
 \hskip -.3cm \,.
\end{equation}

The triangle integral matches the result of BPS form factors, and therefore it captures the full IR divergences. The UV divergence comes solely from the scalar bubble integral ($\sim {1\over \epsilon}$, see \eqref{eq:bubbleintegral}), which is to be cancelled by the renormalization constant: 
\begin{equation}
({\cal Z}^{(1)})_{\Phi_I \Phi_J}^{\Phi_L \Phi_K}  =  -{1\over\epsilon} \left( - \mathds{1} + \mathds{P}  - {1\over 2} \mathds{T} \right)\,. 
\end{equation}
Using \eqref{eq:DvsZ}, we obtain the anomalous dimension density:
\begin{equation}
(\gamma_{\rm SO(6)}^{(1)})_{\Phi_I \Phi_J}^{\Phi_L \Phi_K}  =  2(\mathds{1}- \mathds{P})+ \mathds{T} \,. 
\end{equation}
This is consistent with the dilatation operator \eqref{eq:Hso6} obtained in \cite{Minahan:2002ve}, which can be interperated as a Heisenberg spin chain Hamiltonian.

Reader who is familiar with the computation in \cite{Minahan:2002ve} may note that the above computation based on form factors is quite different. In particular, here one does not need to consider the self-energy contribution. This is because the external states are are on-shell massless states and the massless bubble is zero in dimensional regularization. 
On the other hand, one may also notice that the on-shell condition causes extra complication by introducing IR divergences. At the one-loop, the IR divergence is simply separated in the scalar triangle. For high loops, the IR and UV divergences are in general mixed with each other, which seems to make it non-trivial to extract UV divergences. Fortunately, this problem is easy to solve, thanks to the universality of IR structure. We will discuss this more in Section~\ref{sec:highloopStructure}.

%%%%%%%%%%%%%%%%%%%%%%%%%%%%%%%%
\subsection{SL(2) sector at one-loop}
In this subsection we consider another sector of ${\cal N}=4$ SYM, the SL(2) sector. The operators in this sector contain of a single type of scalar field (denoted as $X$) plus arbitrary powers of covariant derivatives:
\begin{equation}
{\rm tr}(\ldots {D_+^{n_1}X \over n_1!} {D_+^{n_2} X \over n_2!}  {D_+^{n_3} X \over n_3!} \ldots) \,.
\end{equation} 

As in the SO(6) sector, it is enough to consider the density form factor involving two adjacent fields, for which we choose ${{\cal O}_{n_1n_2}}\sim {D_+^{n_1}X \over n_1!} {D_+^{n_2} X \over n_2!}$. Using the double cut in Figure~\ref{fig:F2pt2cutsSO6SL2}, the cut form factor density is given as
\begin{equation}
F_{{\cal O}_{n_1n_2}}^{(1)}(p_i,p_{i+1}) |_{\rm cut} =  \int d \textrm{PS}_2 \,  F_{{\cal O}_{n_1n_2}}^{(0)}(l_1, l_2) \times A_4(p_i, p_{i+1}, l_2, l_1) \,.
\label{eq:FcutSL2}
\end{equation}
The tree minimal form factor is given by simply replacing $D_+$ to be the corresponding on-shell momentum $p_+$ associated to the field $X$ on which the $D_+$ acts. The tree form factor building block is given by 
\begin{equation}
F_{{\cal O}_{n_1n_2}}^{(0)}(l_1, l_2)  = {(l_1^+)^{n_1} \over n_1!} {(l_2^+)^{n_2} \over n_2!}  \,,
\end{equation}
while the tree amplitude is
\begin{equation}
A_4^{(0)}(p_i, p_{i+1}, l_2, l_1)  = \frac{\langle i \, i\textrm{+}1 \rangle \langle l_2 l_1 \rangle}{\langle i\textrm{+}1 \, l_2 \rangle \langle l_1 i \rangle} \,.
\end{equation}
Plugging these tree results in \eqref{eq:FcutSL2}, the cut integrand can be reorganized as
\begin{align}
F_{{\cal O}_{n_1n_2}}^{(1)}(p_i,p_{i+1}) |_{\rm cut} & =-s_{i\,i+1} \frac{(l_1^+)^{n_1}}{n_1!} \frac{(l_2^+)^{n_2}}{n_2!} \begin{tabular}{c}{\includegraphics[height=1.5cm]{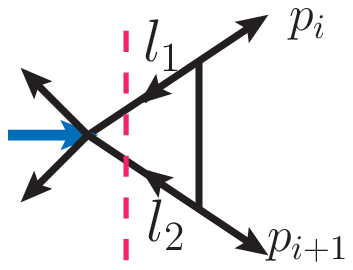} } \end{tabular} \hskip -.5cm \,.
\end{align}
After removing the cuts, one obtains a one-loop triangle integral with the numerator depending on loop momenta $l_i$. 
One can apply tensor integral reduction (see \cite{Loebbert:2016xkw}) to express $l_{i}$ in terms of $\{ p_{i},p_{i+1} \}$ which gives
\begin{equation}
F_{{\cal O}_{n_1n_2}}^{(1)}(p_i,p_{i+1}) = \sum_{m_1,m_2} (I_i^{(1)})_{{n_1}, {n_2}}^{{m_1}, {m_2}} \, {(p_i^+)^{m_1}\over m_1!} {(p_{i+1}^+)^{m_2} \over m_2!} \,, 
\label{eq:FSL21loop}
\end{equation}
where
\begin{align}
(I_i^{(1)})_{{n_1}, {n_2}}^{{m_1}, {m_2}}  =
\left( - s_{i i+1} \delta_{n_1}^{m_1}\delta_{n_2}^{m_2} \right) 
\hskip -.3cm
\begin{tabular}{c}{\includegraphics[height=1.5cm]{scalartriangleSO6SL2_nolabel} } \end{tabular}
\hskip -.6cm 
+ \ (B^{(1)})_{n_1,n_2}^{m_1,m_2} 
 \begin{tabular}{c}{\includegraphics[height=1.5cm]{scalarbubbleSO6SL2} }\end{tabular}  \hskip -.3cm \,.
\end{align}
The coefficient of the scalar bubble integral is:
\begin{equation}
(B^{(1)})_{n_1,n_2}^{m_1,m_2}= - \left(S_{n_1}^{(1)}\delta_{n_1}^{m_1}\delta_{n_2}^{m_2}-
\frac{\theta_{n_2m_2}}{n_2-m_2}\delta_{n_1+n_2}^{m_1+m_2}
+
\left\{\begin{smallmatrix}
n_1\leftrightarrow n_2 \\
m_1\leftrightarrow m_2
\end{smallmatrix}
\right\}\right) 
+ {\cal O}(\epsilon) \,,
\end{equation}
where the harmonic number $S_n^{(\ell)}$ and the Heaviside function $\theta_{nm}$ are defined by
\begin{equation}
S_n^{(\ell)}=\sum_{k=1}^n \frac{1}{k^\ell} \,,
\qquad \qquad
\theta_{nm}=
\left\{\begin{array}{lr}
        1 & \text{for } n>m \,,\\
        0 & \text{for } n\leq m \,.
        \end{array}\right. 
\end{equation}

We can rewrite the $p$ factors in \eqref{eq:FSL21loop} as a minimal tree form factor of ${\cal O}_{m_1,m_2}$, which gives
\begin{equation}
F_{{\cal O}_{n_1n_2}}^{(1)}(p_i,p_{i+1}) =  \sum_{m_1,m_2} (I_i^{(1)})_{{n_1}, {n_2}}^{{m_1}, {m_2}} \cdot F_{{\cal O}_{m_1 m_2}}^{(0)}(p_i,p_{i+1}) \,, 
\end{equation} 
This shows that there is an operator mixing between ${\cal O}_{n_1,n_2}$ and ${\cal O}_{m_1,m_2}$.
The UV divergence comes only from the scalar bubble integrals, and the anomalous dimension density is given by $(-2)$ timing the coefficient of the bubble integral, which is 
\begin{equation}
(\gamma_{\rm SL(2)}^{(1)})_{n_1,n_2}^{m_1,m_2}  = 2\left(S_{n_1}^{(1)}\delta_{n_1}^{m_1}\delta_{n_2}^{m_2}-
\frac{\theta_{n_2m_2}}{n_2-m_2}\delta_{n_1+n_2}^{m_1+m_2}
+
\left\{\begin{smallmatrix}
n_1\leftrightarrow n_2 \\
m_1\leftrightarrow m_2
\end{smallmatrix}
\right\}\right)  \,.
\end{equation}
This is consistent with the result in \cite{Beisert:2003jj}.

%%%%%%%%%%%%%%%%%%%%%%%%%%%%%%%%
\subsection{Structure of high loops}
\label{sec:highloopStructure}

The above procedure can be carried out at higher loops.
First, via unitarity one can obtain the bare form factors. Second, one can renormalize bare form factors by introducing renormalization constants, from which the anomalous dimensions can be obtained. However, there is one complication at high loops which we now explain.

In the one-loop case, there is a clear separation of divergences: the IR divergences come from scalar triangle integral, while UV divergences are determined by the bubble integral.
At higher loop orders, the UV and IR divergences in general can not be separated at integral level in a simple way. 
This problem can be solved by the fact that the IR divergences only depend on the on-shell external particles, which are universal and well understood.\footnote{The IR divergences can be also computed using BPS operators whose form factors are UV free.}

To be concrete, for a renormalized planar form factor $F_{{\cal O}_I}^{\rm ren} = \sum_{J} ({\cal I}^{\textrm{ren}})_I^{~J} F_{{\cal O}_J}^{(0)}$,  its IR divergences take the universal form (which is the same as for planar amplitudes \cite{Bern:2005iz}):
\begin{equation}
\log {\cal I}^{\textrm{ren}} = - \sum_{l=1}^\infty g^{2 l} \bigg[ \frac{\gamma_{\textrm{cusp}}^{({l})} }{(2 {l} \epsilon)^2} + \frac{{\cal G}_{\textrm{coll}}^{({l})} }{2 {l} \epsilon}  \bigg] \sum_{i=1}^n (-s_{i i+1})^{-l \epsilon} \cdot \mathds{1} + {\mathcal O}(\epsilon^0) \,.
\label{eq:IRstructureFF}
\end{equation}
The identity matrix $\mathds{1} = \delta_I^{~J}$ implies that  there is no mixing of IR divergences between different operators.
At two-loop order, it is convenient to use the following form \cite{Bern:2005iz}:
\begin{equation}
(\log {\cal I}^{\textrm{ren}})^{(2)} = {\cal I}^{(2), \textrm{ren}} - {1\over2} \big( {\cal I}^{(1), \textrm{ren}}(\epsilon)\big)^2 = f^{(2)}(\epsilon) {\cal I}^{(1), \textrm{ren}}(2\epsilon) + {\cal R}^{(2)} + {\cal O}(\epsilon) \,,
\label{eq:BDSsubtraction}
\end{equation}
where
\begin{equation}
f^{(2)}(\epsilon) = -2 \zeta_2 - 2\zeta_3 \epsilon - 2 \zeta_4 \epsilon^2 \,,
\end{equation}
and the finite part ${\cal R}^{(2)}$ is usually called the remainder function.
One can check that \eqref{eq:BDSsubtraction} is consistent with \eqref{eq:IRstructureFF} using the cusp and collinear anomalous dimensions given in \eqref{eq:1loopcuspcoll}-\eqref{eq:2loopcuspcoll}.

Therefore, by subtracting the universal IR divergences, one can compute the UV divergences unambiguously.
Below we provide some further details in the SL(2) sector  \cite{Loebbert:2016xkw}.
For the simplicity of notation, we will denote the bare and renormalized loop corrections as
\begin{equation}
{\cal I}^{(l), \textrm{bare}} := {\cal I}^{(l)} \,, \qquad {\cal I}^{(l), \textrm{ren}} := \underline{\cal I}^{(l)}  \,.
\end{equation}

%%%%%%%%%%%%%%%%%%%%
\subsubsection*{SL(2) two-loop case}

In the first step, one needs to obtain bare form factors.
As in the one-loop case, we only need to consider density form factors.
The two-loop correction contains interactions involving both two fields (range-2 interaction) and three fields (range-3 interactions). We denote the corresponding two-loop density corrections as
\begin{align}
& F_{{\cal O}_{n_1n_2}}^{(2)}(p_i,p_{i+1}) =  \sum_{m_1,m_2} (I_i^{(2)})_{{n_1}, {n_2}}^{{m_1}, {m_2}} \cdot F_{{\cal O}_{m_1 m_2}}^{(0)}(p_i,p_{i+1}) \,, \\
& F_{{\cal O}_{n_1n_2 n_3}}^{(2)}(p_i,p_{i+1},p_{i+2}) =  \sum_{m_1,m_2,m_3} (I_i^{(2)})_{{n_1}, {n_2},n_3}^{{m_1}, {m_2}, m_3} \cdot F_{{\cal O}_{m_1 m_2 m_3}}^{(0)}(p_i,p_{i+1},p_{i+2}) \,,
\end{align}
where the first line is the range-2 contribution and the second line is range-3 part. The notation we use is similar to the one-loop case in \eqref{eq:FSL21loop}. The loop corrections $(I_i^{(2)})_{{n_1}, {n_2}}^{{m_1}, {m_2}}$ and $(I_i^{(2)})_{{n_1}, {n_2},n_3}^{{m_1}, {m_2}, m_3}$ can be computed by unitarity cut methods. We will not go into details of this computation but refer interested reader to \cite{Loebbert:2016xkw} for the results. 

\begin{figure}[t]
\centerline{ \includegraphics[height=4.1cm]{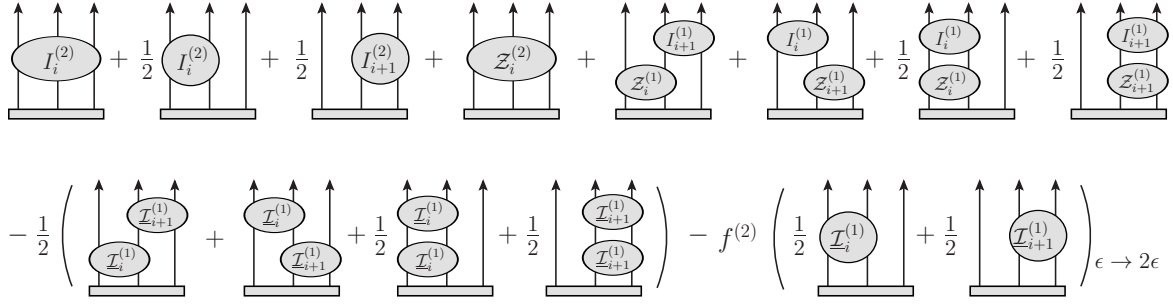} }
\caption{This figure provides the combination of various contributions that give the two-loop density of finite remainder function.}
\label{fig:remainder}
\end{figure}

The IR subtraction and renormalization can be done also at density level. 
Using \eqref{eq:BDSsubtraction}, we express the finite remainder in terms of the renormalized quantities as
\begin{align}
({\cal R}_i^{(2)})_{{n_1}, {n_2},n_3}^{{m_1}, {m_2}, m_3} =  & \ (\underline{\cal I}_i^{(2)})_{{n_1}, {n_2},n_3}^{{m_1}, {m_2}, m_3} 
- {1\over2} \big[ (\underline{\cal I}_i^{(1)})^2 \big]_{{n_1}, {n_2},n_3}^{{m_1}, {m_2}, m_3} \nonumber \\ 
& \ -  f^{(2)}(\epsilon) {1\over2} \bigg[  \delta_{n_3}^{m_3}  ({\underline {\cal I}}_i^{(1)})_{n_1,n_2}^{m_1,m_2} + \delta_{n_1}^{m_1} ({\underline {\cal I}}_{i+1}^{(1)})_{n_2,n_3}^{m_2,m_3} \bigg]_{\epsilon \rightarrow 2\epsilon} \,. 
\end{align}
This can be expanded in various building blocks, which are represented in Figure~\ref{fig:remainder}.

%%%%%%%%%%%%%%
%%%%%%%%%%%%%% 

Let us explain Figure~\ref{fig:remainder} in more details.  The first line of Figure~\ref{fig:remainder} corresponds to the renormalized two-loop correction $(\underline{\cal I}_i^{(2)})_{{n_1}, {n_2},n_3}^{{m_1}, {m_2}, m_3}$ (using \eqref{eq:Fren2loop}):
\begin{align}
&  (\underline{\cal I}_i^{(2)})_{{n_1}, {n_2},n_3}^{{m_1}, {m_2}, m_3} = (I_i^{(2)})_{{n_1}, {n_2},n_3}^{{m_1}, {m_2}, m_3} + {1\over2} \big[ (I_i^{(2)})_{{n_1}, {n_2}}^{{m_1}, {m_2}} + (I_{i+1}^{(2)})_{{n_2}, {n_3}}^{{m_2}, {m_3}} \big] + ({\cal Z}_i^{(2)})_{{n_1}, {n_2},n_3}^{{m_1}, {m_2}, m_3} \\
& +  ({\cal Z}^{(1)})_{n_1,n_2}^{m_1, n_1+n_2-m_1} (I_{i+1}^{(1)})_{n_1+n_2-m_1,n_3}^{m_2, m_3} + ({\cal Z}^{(1)})_{n_2,n_3}^{n_2+n_3-m_3, m_3} (I_{i}^{(1)})_{n_1, n_2+n_3-m_3}^{m_1,m_2} \nonumber\\
& + {1\over2} \bigg[ \delta_{n_3}^{m_3} \sum_k ({\cal Z}^{(1)})_{n_1,n_2}^{k, n_1+n_2-k} (I_{i}^{(1)})_{k, n_1+n_2-k}^{m_1,m_2} + \delta_{n_1}^{m_1} \sum_k ({\cal Z}^{(1)})_{n_2,n_3}^{k, n_2+n_3-k} (I_{i+1}^{(1)})_{k, n_2+n_3-k}^{m_2,m_3} \bigg]  \,. \nonumber
\end{align}
The square of one-loop contribution is
\begin{align}
&  \big[ (\underline{\cal I}_i^{(1)})^2 \big]_{{n_1}, {n_2},n_3}^{{m_1}, {m_2}, m_3} = \\
& ({\underline {\cal I}}_i^{(1)})_{n_1,n_2}^{m_1, n_1+n_2-m_1} ({\underline {\cal I}}_{i+1}^{(1)})_{n_1+n_2-m_1,n_3}^{m_2, m_3} + ({\underline {\cal I}}_{i+1}^{(1)})_{n_2,n_3}^{n_2+n_3-m_3, m_3} ({\underline {\cal I}}_{i}^{(1)})_{n_1, n_2+n_3-m_3}^{m_1,m_2} \nonumber\\
& + {1\over2} \bigg[ \delta_{n_3}^{m_3} \sum_k ({\underline {\cal I}}_i^{(1)})_{n_1,n_2}^{k, n_1+n_2-k} ({\underline {\cal I}}_{i}^{(1)})_{k, n_1+n_2-k}^{m_1,m_2} + \delta_{n_1}^{m_1} \sum_k ({\underline {\cal I}}_{i+1}^{(1)})_{n_2,n_3}^{k, n_2+n_3-k} ({\underline {\cal I}}_{i+1}^{(1)})_{k, n_2+n_3-k}^{m_2,m_3} \bigg]  \,, \nonumber
\end{align}
which is illustrated by the first term in the second line of Figure~\ref{fig:remainder}. 
The renormalized one-loop correction is simply:
\begin{equation}
(\underline{\cal I}_i^{(1)})_{{n_1}, {n_2}}^{{m_1}, {m_2}} = (I_i^{(1)})_{{n_1}, {n_2}}^{{m_1}, {m_2}}  + ({\cal Z}_i^{(1)})_{{n_1}, {n_2}}^{{m_1}, {m_2}} \,.
\end{equation}

Since the whole combination shown in Figure~\ref{fig:remainder} gives the density of finite remainder function, all divergences should cancel with each other in this combination. Therefore, given the two-loop bare form factor results, together with one-loop results, one can determine the two-loop renormalization constants. We refer reader to \cite{Loebbert:2015ova, Loebbert:2016xkw} for more details of applying this strategy at two loops.

%%%%%%%%%%%%%%%%%%%%%%%%%%%%%%%%%%%%
%%%%%%%%%%%%%%%%%%%%%%%%%%%%%%%%%%%%
\section{Conclusion and outlook}
\label{sec:conclusion}

In this review we describe the on-shell formalism for form factors in ${\cal N}=4$ SYM. At tree-level, the minimal form factors can be used to translate off-shell local operators into on-shell language with the help of super spinor helicity formalism.
At loop level, we apply the unitarity method and the color-kinematics duality to construct high loop Sudakov form factors. We also discuss form factor of general non-BPS operators.
Based on these computation, we show that form factors can be used to compute (IR) cusp anomalous dimension and (UV) anomalous dimensions of general local operators.
As mentioned in the introduction, there are many other aspects which we do not cover in this review, and we refer interested readers to the original references. 
The on-shell formalism we focus in this review is based on generic principles and is expected to applicable to general gauge theories. It would be interesting to explore these ideas in more general context.
In this respect, form factors have also been studied in ABJM theory \cite{Brandhuber:2013gda, Young:2013hda, Bianchi:2013pfa}. 
There have been applications of on-shell methods for computing the Higgs amplitudes with high dimension operators in Higgs EFT \cite{Brandhuber:2017bkg, Jin:2018fak, Brandhuber:2018xzk, Brandhuber:2018kqb, Jin:2019ile, Jin:2019opr}.
Form factors with six-dimensional spinor helicity formalism have been studied recently in \cite{Huber:2019fea}.
It would be interesting to apply on-shell form factor techniques to generic effective field theory and their renormalizations, see e.g.~\cite{Brivio:2019irc}. 
It would be also interesting to apply color-kinematics duality to form factor of non-BPS operators or in other gauge theories beyond ${\cal N}=4$ SYM. Finally, it would be interesting to apply the on-shell methods to form factors of multiple operators and correlations functions, see e.g.~\cite{Engelund:2012re}.

%%%%%%%%%%%%%%%%%%%%%%%%%%%%%%%%%%%%
\section*{Acknowledgements}
The author is grateful to Rutger Boels, Andreas Brandhuber, Omer Gurdogan, Tobias Huber, Bernd Kniehl, Florian Loebbert, Andreas von Manteuffel, Dhritiman Nandan, Erik Panzer, Robert Schabinger, Christoph Sieg, Bill Spence, Oleg Tarasov, Gabriele Travaglili, and Matthias Wilhelm for collaboration on various related topics.
He has also benefitted from the discussion with Zvi Bern, John J. Carrasco, Lance Dixon, Bo Feng, Qingjun Jin, Henrik Johansson, Gregory Korchemsky, David Kosower, Jianxin Lu, Jian-Ping Ma, Jan Plefka, Radu Roiban, Volker Schomerus, Vladimir Smirnov, Matthias Staudacher and Congkao Wen at various stages of these projects.
He also thanks Ke Ren for her comments which lead to footnote \ref{footnote8} and \ref{footnote9}.
This work is supported in part by the National Natural Science Foundation of China (Grants No. 11822508, 11847612, 11935013),
by the Chinese Academy of Sciences (CAS) Hundred-Talent Program, 
and by the Key Research Program of Frontier Sciences of CAS. 
%

%%%%%%%%%%%%%%%%%%%%%%%%%%%%%%%%%%%%
%%%%%%%%%%%%%%%%%%%%%%%%%%%%%%%%%%%%
\appendix

%%%%%%%%%%%%%%%%%%%
\section{Feynman rules}
\label{app:FeynRules}

In this appendix, we provide the Feynman rules (in the Feynman gauge) that are used in the main text.  All momenta are taken outgoing.

(1) Feynman rules for the gluon  propagator and gluon vertices with full color dependence are:
\begin{align}
\begin{tabular}{c}{\includegraphics[height=.65cm]{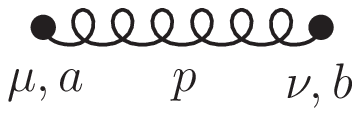} } \end{tabular}
\hskip -.4cm 
= & \ - i  \eta^{\mu\nu}\delta^{ab}{1  \over p^2 + i \epsilon} \,, \\
\begin{tabular}{c}{\includegraphics[height=2cm]{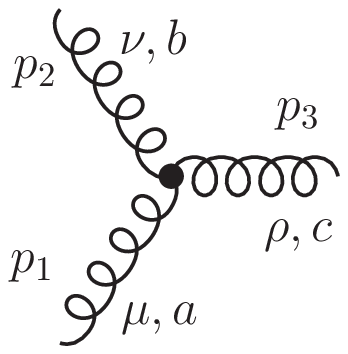} } \end{tabular}
\hskip -.4cm 
= & \  f^{abc} \Big[ \eta^{\mu\nu} (p_1 - p_2)^\rho +  \eta^{\nu\rho} (p_2 - p_3)^\mu +  \eta^{\rho\mu} (p_3 - p_1)^\nu \Big] \,, \\
\begin{tabular}{c}{\includegraphics[height=1.8cm]{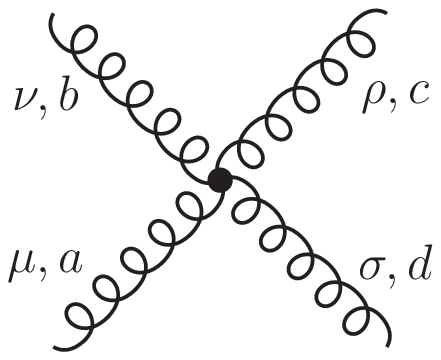} } \end{tabular}
\hskip -.4cm 
= & \ -i f^{abe} f^{cde} \big( \eta^{\mu\rho} \eta^{\nu\sigma} - \eta^{\mu\sigma} \eta^{\nu\rho} \big) -i f^{ace} f^{bde} \big( \eta^{\mu\nu} \eta^{\rho\sigma} - \eta^{\mu\sigma} \eta^{\nu\rho} \big)
\nonumber\\
&  \ -i f^{ade} f^{bce} \big( \eta^{\mu\nu} \eta^{\rho\sigma} - \eta^{\mu\rho} \eta^{\nu\sigma} \big) \,.
\end{align}

(2) Color-stripped Feynman rules which are used to compute color-ordered gluon amplitudes are:
\begin{align}
\begin{tabular}{c}{\includegraphics[height=.65cm]{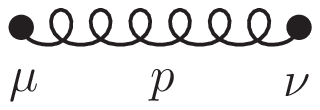} } \end{tabular}
\hskip -.4cm 
= & \ -  i  \eta^{\mu\nu}{1  \over p^2 + i\epsilon} \,, \\
\begin{tabular}{c}{\includegraphics[height=2cm]{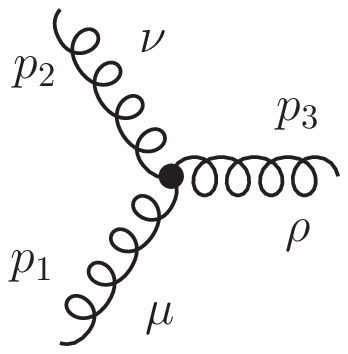} } \end{tabular}
\hskip -.4cm 
= &\  {i \over \sqrt{2}} \Big[ \eta^{\mu\nu} (p_1 - p_2)^\rho +  \eta^{\nu\rho} (p_2 - p_3)^\mu +  \eta^{\rho\mu} (p_3 - p_1)^\nu \Big] \,, \\
\begin{tabular}{c}{\includegraphics[height=1.8cm]{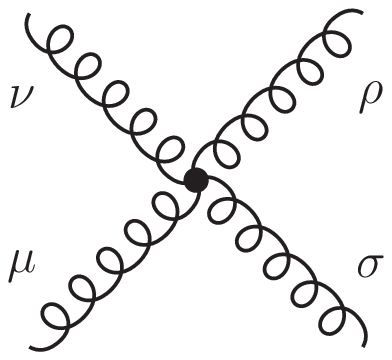} } \end{tabular}
\hskip -.4cm 
= &\  i \eta^{\mu\rho}\eta^{\nu\sigma} - {i\over2} \big( \eta^{\mu\nu}\eta^{\rho\sigma} + \eta^{\mu\sigma}\eta^{\nu\rho} \big) \,.
\end{align}

(3) Color-stripped Feynman rules for the scalar propagator and gluon-scalar interaction vertices are:
\begin{align}
\begin{tabular}{c}{\includegraphics[height=.65cm]{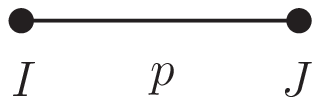} } \end{tabular}
\hskip -.4cm 
= & \ - i \delta_{IJ} {1  \over p^2 + i\epsilon} \,, \\
\begin{tabular}{c}{\includegraphics[height=1.9cm]{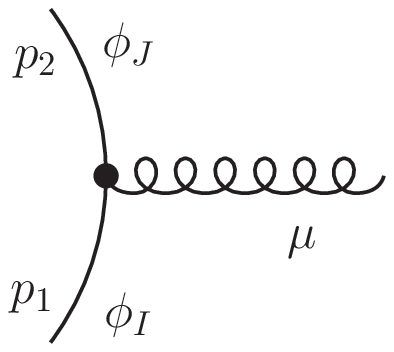} } \end{tabular}
\hskip -.4cm 
= & \ {i\over\sqrt{2}}  \delta_{IJ}(p_1 - p_2)^\mu \,, \\
\begin{tabular}{c}{\includegraphics[height=1.9cm]{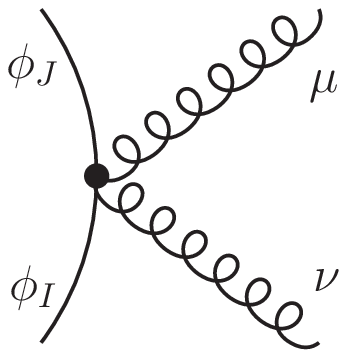} } \end{tabular}
\hskip -.4cm 
= & \ -{i\over{2}}  \delta_{IJ}\eta^{\mu\nu} \,, \\
\begin{tabular}{c}{\includegraphics[height=1.7cm]{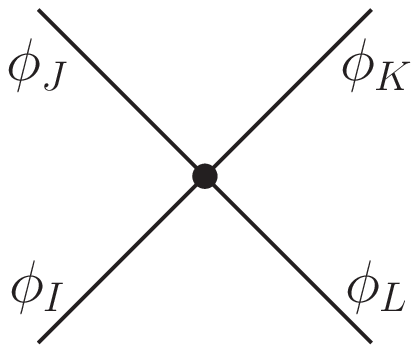} } \end{tabular}
\hskip -.4cm 
= & \ i \delta_{IK}\delta_{JL}  -{i \over 2} \big( \delta_{IJ}\delta_{KL} +  \delta_{IL}\delta_{JK} \big) \,.
\end{align}
%

%%%%%%%%%%%%%%%%%%%
\section{Color algebra}
\label{app:coloralgebra}

Using the completeness relation \eqref{eq:colorcontra} for the generators of gauge group $SU(N_c)$
\begin{equation}
\sum_{a=1}^{N_c^2-1} (T^a)_i^{~j} (T^a)_k^{~l} = \delta_i^{~l} \delta_k^{~j} - {1\over N_c} \delta_i^{~j} \delta_k^{~l} \,,
\end{equation}
it is straightforward to obtain the following relations which are more convenient to use in practice:
\begin{align}
\sum_a {\rm tr}(X T^a Y) {\rm tr}(W T^a Z) & = {\rm tr}(Y X Z W) - {1\over N_c}{\rm tr}(YX) {\rm tr}(ZW)\,, \\
\sum_a {\rm tr}(X T^a Y T^a Z) & = {\rm tr}(ZX) {\rm tr}(Y) - {1\over N_c}{\rm tr}(XYZ) \,,
\end{align}
where $X,Y,Z,W$ can be any product of $T^a$'s.

With the help of the above relations, one can simplify the products of color factors, such as $f^{abc} = {1\over i\sqrt{2}} {\rm tr}([T^a, T^b] T^c)$, to a trace basis. 
As a non-trivial example, let us consider a cube graph. If we assign each vertex a $f^{abc}$ factor and contract all color indices, the color factor is
\begin{equation}
\textrm{color factor of}\Big(\begin{tabular}{c}{\includegraphics[height=1.2cm]{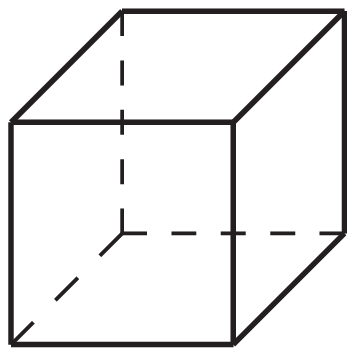} }\end{tabular}\Big) = (f^{\alpha a \beta} f^{\beta b \gamma} f^{\gamma c \delta} f^{\delta d \alpha})^2 = {1\over8}(N_c^2 - 1) N_c^2 (N_c^2 + 12) \,.
\end{equation}
This contains the quatic Casimir group invariants  discussed in Section \ref{sec:IRstructure}.

%%%%%%%%%%%%%%%%%%%%%
\section{Basis of spinor products}
\label{app:spinorbasis}

Using spinor helicity formalism, amplitudes and form factors can be expressed in terms of spinor products $\langle i\, j\rangle$, $[i\,j]$. 
The spinor products satisfy non-trivial non-linear relations, such as the Schouten identity \eqref{eq:schouten}.
This makes it in general hard to compare two expressions. One method is to choose random numerical values for the momenta and then compare expressions numerically. Another way is to use an independent basis of spinor products which we describe in this appendix. Once the results are reduced in the basis, one can compare two expressions analytically. Without loss of generality, we will take five-point amplitudes as an explicit example.  

First, using the relation
\begin{equation}
\vev{i\,i} = 0 =[j\,j] \,, \qquad \vev{i\,j} = -\vev{j\,i} \,, \qquad  [i\,j] = - [j\,i] \,, 
\end{equation}
one only needs to consider
\begin{equation} 
\{\vev{i\,j},  ~ [i\,j]  ~\big|~ i<j \}. 
\label{vev-1} 
\end{equation}

Next, since spinor is two dimensional, it can be expanded in a basis of two spinors. For example, we can use  $\lambda_4, \lambda_5$ as the basis for $\lambda_i$, and $\tilde\lambda_1, \tilde\lambda_2$ as the basis for $\tilde\lambda_i$:
\begin{equation}  
\lambda_i = {\vev{i\,5}\lambda_4-\vev{i\,4}\lambda_5 \over
\vev{4\,5}} \,, \qquad \W\lambda_i =
{[i\,2]\W\lambda_1-[i\,1]\W\lambda_2 \over [1\,2]} \,. 
\end{equation} 
Consequently, the scalar products satisfy
\begin{equation} 
\vev{i\,j} = {\vev{i\,4}\vev{j\,5}-\vev{i\,5}\vev{j\,4} \over \vev{4\,5}} \,, \quad [i\,j] = {[1\,i][2\,j]-[2\,i][1\,j] \over [1\,2]} \,,
\end{equation}
which are equivalent to the Schouten identities. 
With the above relations, we can express all spinor products using the following ones
\begin{equation} 
\{\vev{4\,5},  \,  [1\,2], \, \vev{i\,4}, \, \vev{i\,5}, \, [1\,j], \, [2\,j] ~\big|~ 1\leq i \leq 3, \, 3\leq j \leq 5 \}.
\label{vev-2} 
\end{equation}

For five-point form factors, this will be the final basis. For five-point amplitudes, however, there are four extra constraints from the momentum conservation condition:
\begin{equation} 
P := \sum_{i=1}^5 p_i = 0 \,,
\end{equation}
which in terms of spinor products give four equations:
\begin{eqnarray}  
0 &=& \langle{4| P |1}] = \vev{4\,2}[2\,1]+\vev{4\,3}[3\,1]+\vev{4\,5}[5\,1] \,, \nonumber\\
0 &=& \langle{4| P |2}] = \vev{4\,1}[1\,2]+\vev{4\,3}[3\,2]+\vev{4\,5}[5\,2] \,, \nonumber\\
0 &=& \langle{5| P |1}] = \vev{5\,2}[2\,1]+\vev{5\,3}[3\,1]+\vev{5\,4}[4\,1] \,, \nonumber\\
0 &=& \langle{5| P |2}] =
\vev{5\,1}[1\,2]+\vev{5\,3}[3\,2]+\vev{5\,4}[4\,2] \,. 
\end{eqnarray}  
We can use them to solve for $\{[1\,5], [2\,5], [1\,4], [2\,4]\}$ as
\begin{eqnarray} 
 ~[1\,5] = {\vev{4\,2}[2\,1]+\vev{4\,3}[3\,1] \over \vev{4\,5}} \,,
&\quad& ~[2\,5] = {\vev{4\,1}[1\,2]+\vev{4\,3}[3\,2] \over \vev{4\,5}} \,,
\nonumber\\ ~[1\,4] = {\vev{5\,2}[2\,1]+\vev{5\,3}[3\,1]\over \vev{5\,4}}
\,,&\quad& ~[2\,4] = {\vev{5\,1}[1\,2]+\vev{5\,3}[3\,2]\over \vev{5\,4}} \,.
\label{vev-3} 
\end{eqnarray} 
Thus the final basis for five-point amplitudes can be chosen as:
\begin{equation}  
\{ \vev{1\,4}, \vev{2\,4}, \vev{3\,4}, \vev{1\,5}, \vev{2\,5},
\vev{3\,5}, \vev{4\,5}, [1\,2], [1\,3], [2\,3] \} \,. 
\end{equation} 
For general $n$-point amplitudes, the dimension of the basis is:
\begin{equation}  
n(n-1)-(n-2)(n-3)-4 = 4n-10 \,. 
\end{equation}  
%

%%%%%%%%%%%%%%%%%%%%
\section{Supersymmetric transformation}
\label{app:susy}

In this appendix we discuss the supersymmetric transformation for both the on-shell states and the off-shell fields in chiral operators.

First, we recall the on-shell superfield
\begin{equation}
\Phi(p,\eta)=g_+(p) +  \eta^A \, \bar\psi_A(p) + {\eta^A\eta^B \over 2!} \, \phi_{AB}(p) + { \epsilon_{ABCD} \eta^A\eta^B\eta^C \over 3!} \, \psi^D(p) + \eta^1\eta^2\eta^3\eta^4 \, g_-(p) \,.
\end{equation}
The supersymmetry charges $Q^{\alpha A}$ act as:
\begin{equation}
[Q^{\alpha A}, \Phi(p,\eta)] = \lambda^\alpha \eta^A \Phi(p,\eta) \,.
\label{eq:QonshellNair}
\end{equation}
By matching the components of $\eta$ expansion on both sides of \eqref{eq:QonshellNair}, one can obtain
\begin{align}
[Q^{\alpha A}, g_+(p)] & = - \lambda^\alpha \psi^A(p) \,, \\
[Q^{\alpha A}, \bar\psi_B(p)] & = - \delta_B^A \lambda^\alpha g_+(p) \,, \\
[Q^{\alpha A}, \phi_{BC}(p)] & = \lambda^\alpha \big[ \delta_B^A \bar\psi_C(p) - \delta_C^A \bar\psi_B(p) \big] \,, \\
[Q^{\alpha A}, \psi^B(p)] & = - \lambda^\alpha \phi^{AB}(p) \,, \\
[Q^{\alpha A}, g_-(p)] & = 0 \,.
\end{align}
This can be used to derived SUSY Ward identities for amplitudes and form factors.

Next we consider the transformation on the fields in the chiral operators. The supersymmetric transformation is \cite{Eden:2011yp}:
\begin{align}
Q_A^\alpha \Phi^{BC} & = i \sqrt{2} \big(\delta_A^B \Psi^{C \alpha} - \delta_A^C \Psi^{B \alpha} \big) \,, \\
Q_A^\alpha \Psi_\beta^B & = \delta_A^B F_\beta^{\alpha} + i g_{\rm YM} [\Phi^{BC}, \Phi_{CA}] \delta_\beta^\alpha \,, \\
Q_A^\alpha F_{\beta\gamma} & = \sqrt{2} \, g_{\rm YM} \big( \delta_\beta^\alpha [\Phi_{AB}, \Psi_\gamma^B ] + \delta_\gamma^\alpha  [\Phi_{AB}, \Psi_\beta^B ] \big) \,.
\end{align}
It is obvious that the operator ${\rm tr}(\Phi^{12}\Phi^{12})$ is half-BPS, since it is annihilated by half of SUSY generators $Q_A^\alpha, A=3,4$.
By acting the other four generators $Q_A^\alpha, A=1,2$ on the operator, one can get 
\begin{equation}
\Big( \prod_{A=1}^2\prod_{\alpha=1}^2 Q_A^\alpha \Big) {\rm tr}(\Phi^{12}\Phi^{12}) = 2 \, {\cal L} \,,
\end{equation}
where ${\cal L}$ is the on-shell chiral Lagrangian given in \eqref{eq:calL}.

%%%%%%%%%%%%%%%%%%%%
\section{Loop integrals}
\label{app:integrals}

We use the convention of an $L$-loop integral as
\begin{equation}
I^{(L)}[N(l_i, p_j)]= e^{L\epsilon\gamma_{\rm E}} \int\frac{d^D l_1}{i\pi^{\frac{D}{2}}}\dots\frac{d^Dl_L}{i\pi^{\frac{D}{2}}}
\frac{N(l_i, p_j)}{\prod_j D_j} \,.
\end{equation}
The dimensional regularization is used with $D=4-2\epsilon$.
Below we list some explicit results of one- and two-loop integrals used in main text. They can be evaluated with standard integration methods \cite{Smirnov:2012gma}. 

The one-loop scalar triangle integral is:
\begin{align}
I_{\rm tri}^{(1)} &= \hskip -.2cm \begin{tabular}{c}{\includegraphics[height=1.2cm]{scalartriangle} } \end{tabular} 
\hskip -.4cm =  {e^{\epsilon\gamma_{\rm E}}} \int\frac{d^D l}{i\pi^{\frac{D}{2}}}
\frac{1}{l^2 (l+p_1)^2 (l+p_1+p_2)^2} \nonumber\\
&= \big[e^{\epsilon\gamma_{\rm E}} (-s_{12}^2)^{-\epsilon}\big] \bigg[ -  (-s_{12}^2)^{-1} \frac{\Gamma(-\epsilon)^2\Gamma(1+\epsilon)}{\Gamma(1-2\epsilon)}  \bigg] \nonumber\\
&=  (-s_{12}^2)^{-\epsilon-1} \Big[ - {1\over\epsilon^{2}} + {\pi^2 \over 12} + {7 \zeta_3 \over3} \epsilon + {47 \pi^4 \over 1440} \epsilon^2  + {\cal O}(\epsilon^3) \Big] \,.
\label{eq:scalartriangleintegral}
\end{align}
The one-loop scalar bubble is:
\begin{equation}
I_{\rm bub}^{(1)} = {-\epsilon \over 1-2\epsilon} I_{\rm tri}^{(1)} 
= (-s_{12}^2)^{-\epsilon} \Big[ {1\over\epsilon} + 2 + \Big(4 - {\pi^2 \over 12} \Big)\epsilon + \Big(8 - {\pi^2 \over 6} - {7\zeta_3\over3} \Big) \epsilon^2  + {\cal O}(\epsilon^3) \Big]
 \,.
 \label{eq:bubbleintegral}
\end{equation}
The two-loop planar-ladder and cross-ladder scalar integrals are
\begin{align}
I^{(2)}_{\rm PL} &= \hskip -.2cm \begin{tabular}{c}{\includegraphics[height=1.2cm]{planarladder} } \end{tabular} 
\hskip -.4cm = (-s_{12}^2)^{-2\epsilon-2} \Big[ {1\over4 \epsilon^{4}} +  {5 \pi^2 \over 24 \epsilon^{2}} + { 29 \zeta_3 \over 6 \epsilon} + {3 \pi^4 \over 32} + {\cal O}(\epsilon^1) \Big] \,, \\
I^{(2)}_{\rm CL} &= \hskip -.2cm \begin{tabular}{c}{\includegraphics[height=1.cm]{crossladder} } \end{tabular} 
\hskip -.4cm = (-s_{12}^2)^{-2\epsilon-2} \Big[ {1\over \epsilon^{4}} - {\pi^2 \over \epsilon^{2}} - { 83 \zeta_3 \over 3 \epsilon} - {59 \pi^4 \over 120} + {\cal O}(\epsilon^1)  \Big]\, . 
\end{align}

%%%%%%%%%%%%%%%%%%%%%%%%%%%%%%%%%%%%
%%%%%%%%%%%%%%%%%%%%%%%%%%%%%%%%%%%%

\providecommand{\href}[2]{#2}\begingroup\raggedright\endgroup

\end{document}